\documentclass[aps,twocolumn,showpacs,amsmath,amssymb]{revtex4}
\usepackage{amsmath}
\usepackage{amssymb}
\usepackage[colorlinks,plainpages=false,linkcolor=blue,urlcolor=blue,citecolor=blue,pdfpagemode=UseNone,pdfstartview=FitH]{hyperref}
\usepackage[T1]{fontenc}
\usepackage[utf8]{inputenc}
\usepackage[english]{babel}
\usepackage{color}
\usepackage{float}
\usepackage{graphicx}
\usepackage{epstopdf}
\usepackage{booktabs}
\usepackage{mathtools}
\usepackage{bigstrut}

\newcommand{\zSe}{$z_\text{Se}$}
\newcommand{\zSeC}{$z_\text{Se}/c$}
\newcommand{\alat}{$a$}
\newcommand{\clat}{$c$}
\newcommand{\ca}{$c/a$}
\newcommand{\ad}{$a$-direction}
\newcommand{\cd}{$c$-direction}

\newcommand{\ai}{\textit{ab initio}}
\newcommand{\intra}{intralayer}
\newcommand{\inter}{interlayer}
\newcommand{\Grimme}{DFT-D2}
\newcommand{\TS}{DFT-TS}
\newcommand{\AAA}{\,\AA$^3$}
\newcommand{\kp}{k-point}

\begin{document}

\title{\textit{Ab initio} investigation of lattice distortions in response to van der Waals interactions in FeSe}

\author{Felix Lochner$^{1,2}$, Ilya M. Eremin$^{2}$, Tilmann Hickel$^{1}$, and J\"org Neugebauer$^{1}$}

\affiliation{$^1$Max-Planck-Institut f\"ur Eisenforschung, D-40237 D\"usseldorf, Germany}

\affiliation{$^2$Institut f\"ur Theoretische Physik III, Ruhr-Universit\"at Bochum, D-44801 Bochum, Germany}

\begin{abstract}
	The electronic structure in unconventional superconductors holds a key to
	understand the momentum-dependent pairing interactions and the resulting superconducting gap function. In superconducting Fe-based chalcogenides, there have been controversial results regarding the importance of the $k_z$ dependence of the electronic dispersion, the gap structure and the pairing mechanisms of iron-based superconductivity. Here, we present a detailed investigation of the van der Waals interaction in FeSe and its interplay with magnetic disorder and real space structural properties. Using density functional theory we show that they need to be taken into account upon investigation of the 3-dimensional effects, including non-trivial topology, of FeSe$_{1-x}$Te$_x$ and FeSe$_{1-x}$S$_x$ systems. In addition, the impact of paramagnetic (PM) disorder is considered within the spin-space average approach. Our calculations show that the PM relaxed structure supports the picture of different competing ordered magnetic states in the nematic regime, yielding magnetic frustration.
\end{abstract}

\date{\today}


\maketitle

\section{Introduction} \label{Sec:Int}
Among  the several types of Fe-based superconductors (FeSC) discovered so far, FeSe has the simplest crystal structure consisting only of superconducting layers. This system turns out to be unique not only due to its structural properties but also due to the lack of any magnetic transition at ambient pressure \cite{Bohmer.2018}. The tetragonal to orthorhombic structural transition at $T_s\approx 90$\,K \cite{McQueen.2009b, Margadonna.2008}, where the orthorhombic structure is called nematic phase analogue to liquid crystals \cite{Fernandes.2012b}, occurs in FeSe without the presence of any ordered magnetic state \cite{McQueen.2009b, Bendele.2010}.  In addition, the interplay of physical pressure and chemical substitution causes dramatic changes in the phase diagram \cite{Medvedev.2009, Bendele.2012, Sun.2016, Kothapalli.2016}.
This underlines the delicate interplay between real space crystal structure and the electronic properties including magnetism in this compound.

Although electronic correlations are relatively strong in Fe-based superconductors, the application of density functional theory (DFT) has proven to reliably provide insights into their physical properties. The band structure calculations correctly predict the main electronic properties for most of the Fe-based superconductors including their Fermi surface topology, structural and magnetic transitions and the possible strength of electron-phonon interactions \cite{Guterding.2017}. In combination with a projection onto Wannier functions, based on symmetry considerations, it further yields low-energy models that are then used to describe the broken symmetries in these systems \cite{Eschrig.2009, Lochner.2017}.

In most of the Fe-based superconductors the magnetic ground state is in DFT and experiment given by the $C$-type anti-ferromagnetic order (often dubbed as stripe-type anti-ferromagnetic (sAFM) order) \cite{Johnston.2010, Hosono.2018, Dai.2015}. For FeSe DFT also predicts the sAFM state to be the ground state \cite{Li.2009b}. However, in experiment this magnetic phase appears only at finite pressure, while at ambient pressure the so-called nematic (structural) transition without any long range magnetic order is observed \cite{Khasanov.2008, Hsu.2008, Margadonna.2008, McQueen.2009b}. Several \ai\ calculations explained the absence of the long-range magnetic order by the presence of competing magnetic phases with different ordering vectors \cite{Liu.2016}. The resulting magnetic frustration prevents the formation of the long-range magnetic order, but allows the nematic transition, which is breaking $Z_2$ symmetry \cite{Glasbrenner.2015, Christensen.2019, Ruiz.2019, Busemeyer.2016}. This mechanism is supported by recent experiments indicating two different types of magnetic fluctuations in FeSe \cite{Baum.2019}.

Especially the complex pressure dependent phase diagram of FeSe is highly debated \cite{Bohmer.2019}. Moreover, a clear impact of the nematic transition on the magnetism in the material indicates a strong coupling between the structure and spin fluctuations \cite{Wang.2016c, Wang.2016d}. To get a deeper understanding of the structural properties in the absence of magnetic order at low temperatures, as it is seen for the nematic state, we avoid long-range magnetism by employing a paramagnetic (PM) approach based on the spin space averaging technique \cite{Walle.2002}.

\begin{figure}[h]
	\centering
	\includegraphics[width=4.5cm]{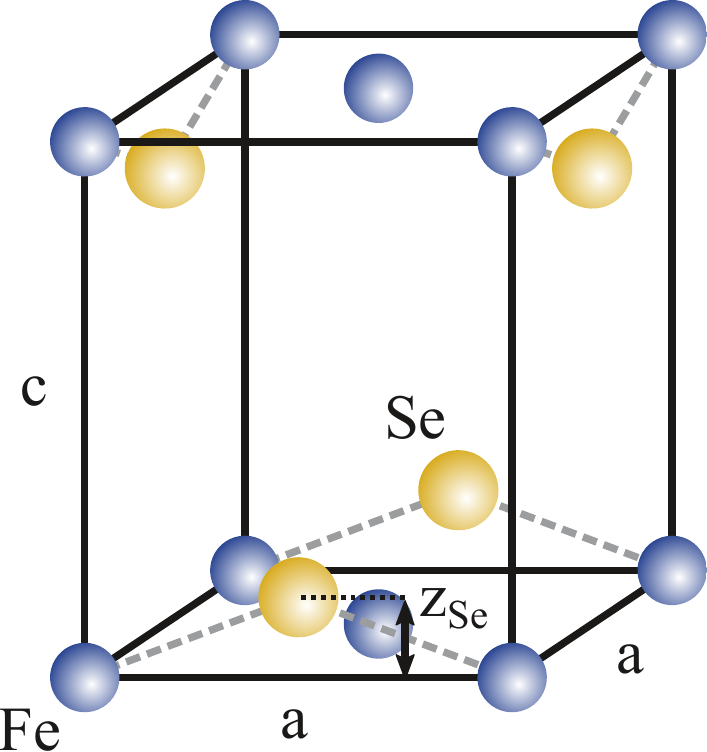}
	\caption{Atomic structure of tetragonal FeSe. The three structural parameters that have to be relaxed are the $a$ and $c$ lattice constants as well as \zSe\,, i.e., the height of the Se-atoms with respect to the iron layer.}
	\label{Fig:FeSe_structure}
\end{figure}

As we know from previous works, superconductivity in FeSe is quite sensitive to structural changes, in particular to the height of the Se-atoms above the iron layer, \zSe\ \cite{Guterding.2017b} (see Fig.~\ref{Fig:FeSe_structure}). More generally, the three dimensional superconductivity and the physics explaining the deviation of calculated lattice parameters from experimental results are of high interest.
In particular there are indications that dynamic dipole-dipole (van der Waals (vdW)) interactions might be important, as they drive \inter\ attraction \cite{Guterding.2017b, Ricci.2013}. The nature of these interactions can only be resolved, if different models for the vdW interactions are compared. Those implementations allow us to investigate the nature of the interactions between atom species in detail \cite{Grimme.2006, Tkatchenko.2009}.  Moreover, the interplay of vdW and PM allows us to access charge based interactions without breaking the translational symmetry.

The paper is organized as follows. In section~\ref{Sec:Methods} we describe the PM and vdW extensions to standard DFT calculations. In section~\ref{Sec:PM} the implementation of paramagnetic DFT is discussed. In section~\ref{Sec:vdW} we compare two different vdW-correction schemes in application to FeSe for the sAFM and the PM state. Finally, we summarize our results in Section~\ref{Sec:Sum}.

\section{Methods} \label{Sec:Methods}
\subsection{Paramagnetism}
In contrast to experiment, standard DFT predicts the sAFM state to be stable in FeSe and is commonly used although other magnetic stats are close in energy. Due to this degeneracy the competition of several magnetic configurations may yield frustration and prevent the stability of long-rang magnetic order in FeSe. It is well known that the presence of magnetic order and disorder has a strong effect on the structural properties of Fe-based materials \cite{Bleskov.2016}. This calls for a systematic study within the PM state to investigate the structural behavior of the system in the nematic phase without long-range magnetic order and with higher precision. In particular, we construct different special quasi-random structures (SQS's) to maximize the magnetic disorder in a finite simulation box. This approach is based on geometrical considerations and determines for a given real space structure the best possible spin configuration to mimic the PM state. For FeSe we distinguish three species (Se-atoms, Fe-up-atoms, Fe-down-atoms), where the SQS is performed for the Fe-atoms. Hereby, we used the ATAT package \cite{Walle.2002} to create three different setups of the magnetic moments. The difference of those SQS's is the real space cutoff radii for the correlations considered in the construction algorithm. To make a systematically treatment of the magnetism possible, the magnetic moments of each Fe-atom has the same magnitude. 

Due to convergence issues with automatized relaxation algorithms for the ionic positions and the magnetic moments, the height of Se-atoms with respect to the iron layer (\zSe), the volume and \ca\ ratio were relaxed sequentially. Moreover constrained magnetic moments are used, where both direction and magnitude are constrained, to avoid non-magnetic (NM) final structures.

To calculate the local magnetic moments an integral over a sphere $\Omega_I$, which depends on a radius given by the chosen structure and the volume of the unit cell (UC) in particular, is taken. This system specific radius is chosen in such a way, that 98\% of the UC is used for evaluating the magnetic moment, what also leads to an overlap of some of those spheres within the UC. 

To get the magnetic moment $m_\text{opt}$ that is energetically preferred by the system, we fit the total energy to a polynomial function 
\begin{align}
f(m) = \alpha m^2+\beta m^4\text{ with } \alpha,\beta\in \mathbb{R}\text{ ,} \label{Eq:Magmom_fit}
\end{align}
which is the simplest form of the Landau free energy expansion.

\subsection{Selected van der Waals implementations}
FeSe is a compound with strongly different polarizabilities, thus the distribution of Fe and Se atoms yields different \inter\ and \intra\ vdW contributions. Therefore we have carefully investigated the interdependence of vdW interactions and modifications of all lattice parameters. In order to resolve this interdependence, we have chosen two different vdW approaches. On the one hand, the \Grimme\ approach of Grimme \cite{Grimme.2006}, referring to dispersion correction version 2, is a conceptually more simple approach. On the other hand, we employ \TS\ of Tkatchenko and Scheffler \cite{Tkatchenko.2009}, which includes the desired geometrical weighting based on the used compound. Those represent different levels of complexity. For FeSe the \Grimme\ method has already been used, a detailed comparison to the \TS\ method, however, is so far missing \cite{Ricci.2013}. While for details about the individual approaches we refer to Refs.~\cite{Grimme.2006, Tkatchenko.2009}, we focus the upcoming discussion on the differences of the \Grimme\ method and the \TS\ method.

In general the vdW interactions are introduced by adding a semi-phenomenological correction $E_\text{vdW}$ to the standard DFT energy given by
\begin{align}
  \label{eq:E_vdW}
E_\text{vdW}=-\frac{s}{2}\sum\limits_{A,B}^{N_\text{at}} f_\text{dmp}\left(R_{AB}\right)\frac{C_{6,AB}}{R_{AB}^6}\text{ ,}
\end{align}
where $s$ is a global scaling factor depending on the chosen exchange-correlation functional and usually obtained by least-square fits for the total energy deviations of different test samples (see \cite{Grimme.2006} for details). $N_\text{at}$ describes the total number of atoms in the unit cell, $A$ and $B$ denote different atoms, where $R_{AB}$ refers to the distance between these atoms. $C_{6,AB}$ is the corresponding $C_6$ parameter of the polarizability and $f_\text{dmp}$ is a global damping function.

The main difference between those two approaches is the choice of the $C_{6,AB}$ parameters. For \Grimme\ least square fits of experimentally found atomic $C_6$ parameters given by the dipole oscillation strength distribution method (DOSD) are taken to generate the $C_6$ parameters for the calculations \cite{Wu.2002}. For binary contributions the geometric mean is taken, $C_{6,AB}=\sqrt{C_{6,A}C_{6,B}}$ \cite{Grimme.2006}. This by construction ignores the specific atomic configuration in a given system and is less suited for bulk materials and alloys. For example, in the case of FeSe the \intra\ Se-Se interaction is calculated without including the underlying Fe-layer.

For the \TS\ method the $C_{6,AA}$ parameters are derived from the parameters for one free standing atom of the same species given by
\begin{align}
C_{6,AA}=\frac{\eta_A}{\eta_A^\text{at}}\left(\frac{\kappa_A}{\kappa_A^\text{at}}\right)^2\left(\frac{V_A}{V_A^\text{at}}\right)^2C_{6,AA}^\text{at}\text{ ,}
\end{align}
where the index "at" marks properties of the free standing atom. All quantities without the index refer to the effective parameters of the full system. The $C_{6,AA}^\text{at}$ are taken from Ref. \cite{Chu.2004} and $\eta$ is an effective frequency (introduced in the London formula \cite{Tang.1969}). Here the effective volume $V_A=\kappa_A\alpha_A$ dresses the $C_6$ parameter by including the local environment in the form of $V_A/V^\text{at}_A$, where the Hirschfeld atomic partitioning weights $w_A(\mathbf{r})$ \cite{Hirschfeld.1977} are used to obtain this ratio. $\kappa_A$ is a scaling factor for the polarizability $\alpha_A$. The ratio is given by
\begin{align}
\frac{\kappa_A}{\kappa_A^\text{at}}\frac{\alpha_A}{\alpha_A^\text{at}}=\frac{V_A}{V_A^\text{at}}=\frac{\int r^3 w_A(\mathbf{r})n(\mathbf{r})\mathrm{d}^3\mathbf{r}}{\int r^3 n_A^\text{at}(\mathbf{r})\mathrm{d}^3\mathbf{r}}\text{ ,} \label{Eq:V_eff}
\end{align}
with
\begin{align}
w_A(\mathbf{r})=\frac{n_A^\text{at}(\mathbf{r})}{\sum_Bn_B^\text{at}(\mathbf{r})}\text{ ,} \label{Eq:Hirschfels_weight}
\end{align}
where the sum over $n_B^\text{at}(\mathbf{r})$ is the so called promolecule electronic density. As a combination rule for $C_{6}$ parameters \TS\ uses the expression
\begin{align}
C_{6,AB}=\frac{2C_{6,AA}C_{6,BB}}{\frac{\alpha_B}{\alpha_A}C_{6,AA}+\frac{\alpha_A}{\alpha_B}C_{6,BB}}\text{ .}
\end{align}

\section{Paramagnetic Calculations} \label{Sec:PM}
The calculations are performed using the Vienna Ab-Initio Simulation Package (VASP) \cite{Kresse.1993, Kresse.1996, Kresse.1996b} with the projector augmented wave method \cite{Blochl.1994} and Perdew, Burke, and Ernzerhof \cite{Perdew.1996} type exchange-correlation functionals. Calculations for the UC use a $12\times 12\times 8$ Monkhorst-Pack \kp s mesh \cite{Monkhorst.1976}. The PM calculations are performed in a  $2\times 2\times 2$ supercell with a $6\times 6\times 4$ \kp\ mesh accordingly. The energy cut-off is set to 450\,eV justified by a convergence of the bulk modulus to an accuracy of 1\%. For the electronic smearing we chose first order Methfessel-Paxton smearing \cite{Methfessel.1989} with $\sigma=0.1$\,eV. Setting up all calculations, postprocessing and analyzing is done by using the integrated development environment pyiron \cite{Janssen.2019}.

\begin{figure}[h]
	\centering
	\includegraphics[width=8.6cm]{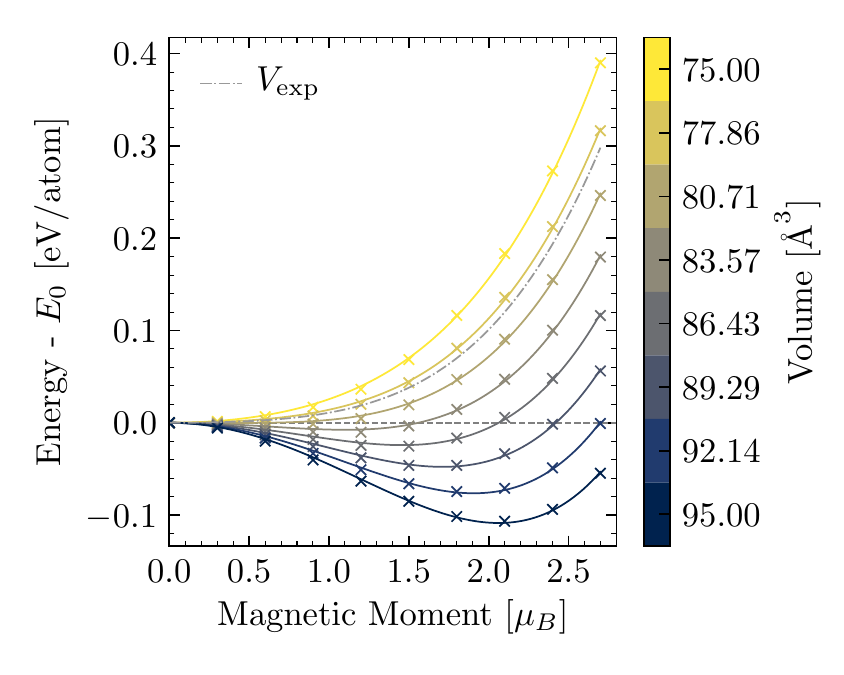}
	\caption{Fe-atom magnetic moment dependent total energy of spin constrained \TS\ calculations aligned to its energy of zero magnetization. The polynomial fits are obtained from Eq.~\eqref{Eq:Magmom_fit}. The colorbar shows to the chosen volumes. The gray dashed dotted line indicates the dependency for the experimental volume at 298 K taken from \cite{McQueen.2009}.}
	\label{Fig:Magmom_pm_cons}
\end{figure}

To investigate the disordered magnetic phase, we constrain the magnetic moments of the Fe-atoms as described in the previous section. To compare the influence of disorder to the structural properties, we also calculate the equilibrium state structural parameters for the sAFM state for comparison. As explained later in the section, the cell shape is in the PM calculations not fully converged. Since the differences in structure and energy for each calculation step are small, we took a snapshot after 9 complete minimization steps for the following analysis. The exemplary results for one used SQS spin setup are illustrated in Fig.~\ref{Fig:Magmom_pm_cons}, where the fits are obtained by Eq.~\eqref{Eq:Magmom_fit}. Here the transition from the NM state at $V=75$\AAA\ to the PM state at $V=95$\AAA\ can be clearly seen. Only by constraining the magnetic moments, the smooth phase transition can be investigated. Extracting the minimal energy values and corresponding magnetic moments, we calculate the equilibrium volume for each SQS.

\begin{figure}[h]
	\centering
	\includegraphics[width=8.6cm]{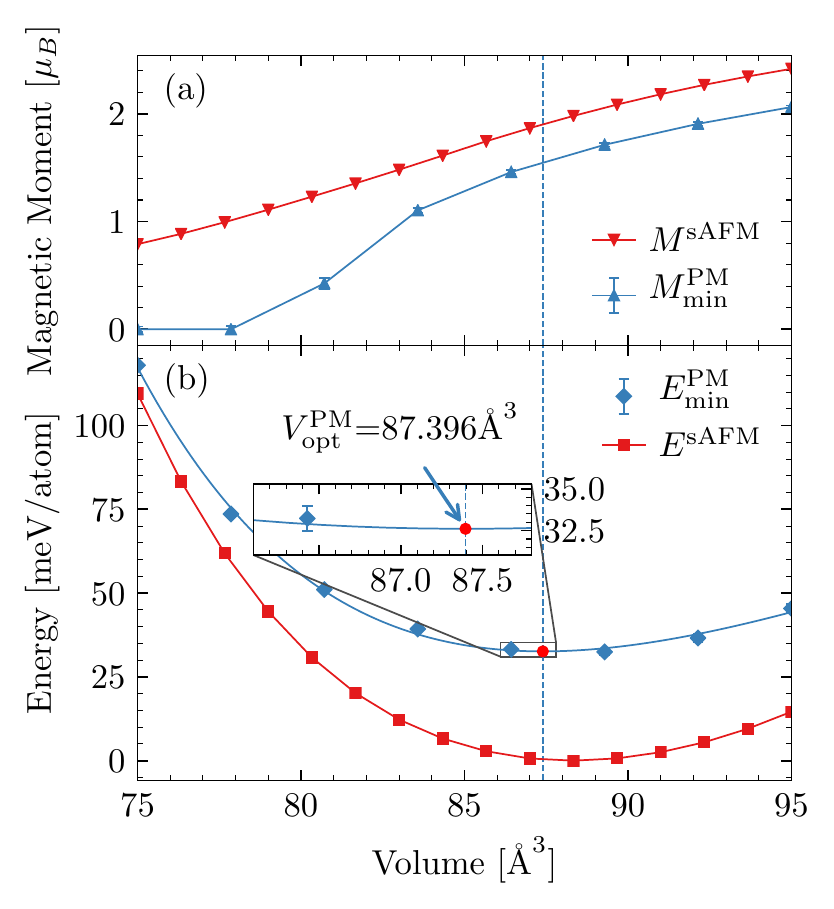}
	\caption{(a) Magnetic moments and (b) total energies  per Fe-atom as obtained at the minima of the fits in Fig.~\ref{Fig:Magmom_pm_cons} compared to the sAFM state (red solid line) for the same structural parameters. The error bars are connected to the fit and the dashed line indicates the optimal volume for the SQS configuration. The optimal volume is obtained by Rose-Vinet equation of states \cite{Vinet.1987} and the fit is shown by the solid curves.}
	\label{Fig:Magmom_Energy_pm_min_all_cons}
\end{figure}

Since several SQS's are required to get sufficient statistics for the PM average, the energy-volumes curves for the mean of three SQS's are shown for one exemplary relaxations step in Fig.~\ref{Fig:Magmom_Energy_pm_min_all_cons}~(b). Since there is almost no scatter between the different structures, the setup for the SQS (i.\,e., the size of the supercell) is sufficient to describe the PM ground state.

Moreover, the obtained fit to the Rose-Vinet equation of states \cite{Vinet.1987} is in good agreement with the given data, confirming a smooth transition from the PM ($V=87$\AAA) to the NM ($V\approx 80$\AAA) ground state. We learn from Fig.~\ref{Fig:Magmom_Energy_pm_min_all_cons}~(a) that the equilibrium state has a finite magnetic moment, while the NM state does not represent an equilibrium state. The larger error bar for the intermediate volumes ($V\approx 81$\AAA) between the  PM and NM state reflects the competition of these phases, when both are close in energy. Those errors, however, do not show a large influence on the Rose-Vinet fit, which means PM calculations are still sufficiently reliable even in the regions of comparably small magnetic moments. Also the mean magnetic moment of the Fe-atoms at the equilibrium lattice constant ($M^\text{PM}_\text{opt} \approx 1.6\,\mu_B$) is comparable for PM and sAFM ($M^\text{sAFM}_\text{opt} \approx 1.8\,\mu_B$). We see a small variation of the magnitude of the magnetic moments for the Fe-atoms in the PM state, what is in the tolerance of the used approach. This might be related to several magnetic sublattices with different absolute magnetic moments, what was discussed for the 11 compounds previously \cite{Gastiasoro.2015}. As the variation is small, we consider this effect to be less important for the structural parameter relaxation. 

It should also be noticed, that the total energy per atom at the equilibrium for the PM state is about $30$\,meV larger than that one for the sAFM state. It indicates that the PM state is not predicted as the ground state in the present DFT approach, but would only become stabilized by magnetic entropy. However, the PM calculations include constrained magnetic moments, thus the total energy includes a penalty which is also in the order of $10$\,meV. Since a competition of several short range magnetic orders is also considered to drive the PM state in the nematic region \cite{Liu.2016}, it might also explain the remaining difference as ordered states are usually lower in energy.

Although the equilibrium volume changes only by less than 1\% for the PM state compared to the that of the sAFM one later discussed for Fig.~\ref{Fig:Structural_param_all}, the structural parameters show a more critical behavior due to magnetically driven changes. In Fig.~\ref{Fig:Structural_param_all} it can be seen that the lattice parameter in \ad\ is significantly reduced while the lattice parameter $c$ increases, what is in contradiction to the experimental results. This is caused by a lack of \inter\ interactions also found by a previous work \cite{Ricci.2013}. The lack of \inter\ attraction is presumably caused by missing vdW interactions in the DFT calculations for FeSe. 

As an interim summary we see that the magnetic moment gives a smooth transition from the NM to the PM state, without any convergence issues for the magnetic moments. By introducing the PM state, the \inter\ interactions are reduced compared to the sAFM state and not able to glue both FeSe-layers together, resulting in a quasi-free behavior of those layers. It also explains that the cell shape in our calculations never converges. We observe both layers drifting apart for each minimization step with no significant change in total energy, caused by missing \inter\ attraction. This indicates that magnetic interactions are not able to explain the \inter\ coupling in this compound for a disordered magnetic state. Since the magnetic moment of the Fe-atoms in the PM state is comparable to that of the sAFM calculations, the covalent bonds are considered to not explain the \inter\ attraction, however, a minimum in the total energy is observed for the sAFM state.

\section{Van der Waals Calculations} \label{Sec:vdW}
To account for the \inter\ attraction mentioned in the previous section, we include vdW forces  for the PM state. In order to decouple it from magnetic disorder and due to numerical efficiency, however, we start the analysis of vdW corrections for the sAFM state. For that part we investigate the structural response for the \Grimme\ and the \TS\ approach, where we analyse the performance of these methods by analyzing the response to volume and shape changes for FeSe.

\subsection{Total Energies}
The sAFM ordered magnetic configuration is known to be the magnetic ground state of many other FeSC's \cite{Bohmer.2018}. Moreover, most of the possible stable anti-ferromagnetic structures are energetically close and their impact on the atomic structures is similar \cite{Liu.2016}. This makes our anti-ferromagnetic calculations representative for Fe-based chalcogenides.

In DFT and \ai\ thermodynamics it is required to analyze the energetics of materials as a function of volume. Here, we particularly use this dependence to investigate the origin of the electronic \inter\ and \intra\ interactions. If both interactions are of the same origin and magnitude, the ratio between the lattice constants \alat\ and \clat\ is expected to change only weakly after a cell-shape relaxation. Moreover, \zSe\ should be proportional to changes in the volume.

For the volume dependent calculations we relax the cell shape as well as the ionic positions simultaneously. Although the full relaxation may lead in special cases to slightly different positions and energy values due to changing numbers of plane waves in the calculation, we do not find any significant changes of the structure compared to relaxing one quantity after another.
\begin{figure}[h]
	\centering
	\includegraphics[width=8.6cm]{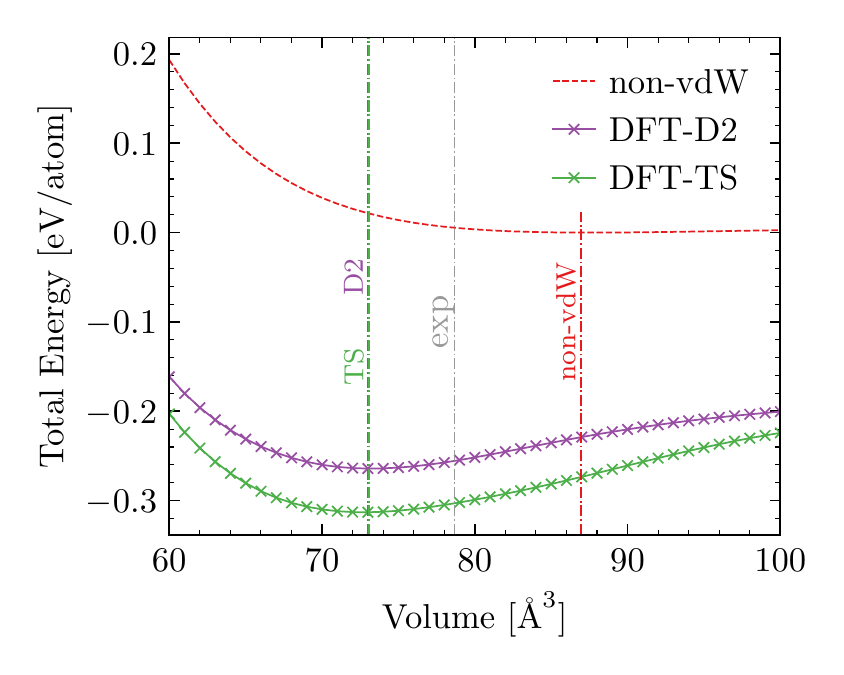}
	\caption{Calculated total energy vs. volume for the non-vdW sAFM state (red dashed) compared to vdW corrected calculation by the \Grimme\ method (violet solid) and \TS\ one (green solid). The values are aligned to the minimum of the total energy of the non-vdW curve. The dashed dotted lines indicate the experimental volume at 298 K by \cite{McQueen.2009} (grey), the non-vdW optimized volume (red) and the optimized ones for the \Grimme\ approach (violet) and the \TS\ approach (green).}
	\label{Fig:AFM_energy_vdW}
\end{figure}

\begin{figure*}[htbp]
	\centering
	\includegraphics[width=17.2cm]{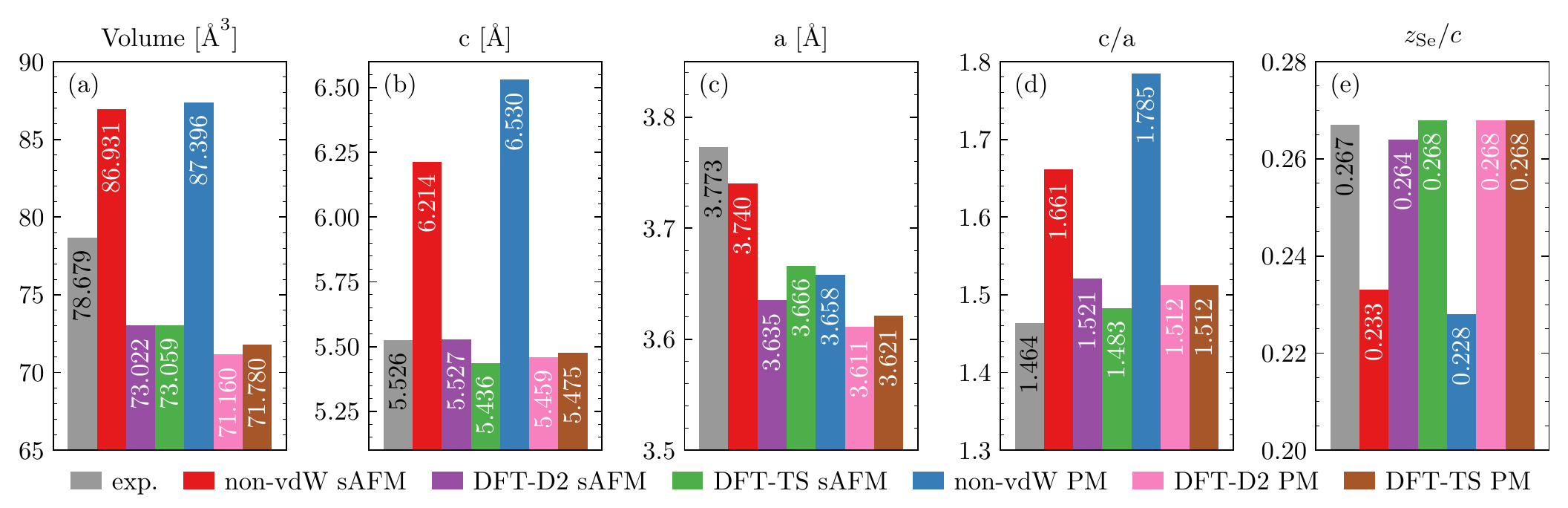}
	\caption{Lattice parameters for FeSe in the sAFM and PM state with and without vdW corrections. The experimental values by \cite{McQueen.2009} (gray), the numerical values for non-vdW sAFM calculations (red), sAFM \Grimme\ calculations (violet), sAFM \TS\ calculations (green), non-vdW PM calculations (blue), PM \Grimme\ calculations (pink) and PM \TS\ calculations (brown) are illustrated for (a) the minimum volume of the UC, (b) the lattice parameter in \cd, (c) the lattice parameter in \ad, (d) the fraction of lattice parameter $c$ and $a$ and (e) the height of the Se-atom \zSe\ in units of $c$.}
	\label{Fig:Structural_param_all}
\end{figure*}

The total energy behavior in Fig.~\ref{Fig:AFM_energy_vdW} shows a shallow minimum for non-vdW sAFM. The small binding energy indicates  that the material could be unstable already at moderate temperatures. In contrast, both \Grimme\ and \TS\ show a clear minimum at $V_\text{min}^\text{D2}=73.022$\AAA\ and $V_\text{min}^\text{TS}=73.059$\AAA. The corresponding volumes are smaller than the non-vdW optimized volume at $V_\text{min}^\text{non-vdW}=86.931$\AAA\ which is about $\approx 10$\AAA\ larger than the experimentally measured volume \cite{McQueen.2009} (see Fig.~\ref{Fig:Structural_param_all}~(a)). 
We will discuss below to what degree the nearly similar optimal volume for \Grimme\ and \TS\ is coincidence or indicating the similarity of the implemented physical concepts. For non-vdW calculations the deviation of the lattice constant \alat\  from the experimental value is only about 1\%, yet the lattice parameter \clat\ is $ \sim 13\%$ too large. This indicates that the failure of the uncorrected DFT Hamiltonian primarily affects the \inter\ interactions. As discussed below the smaller vdW corrected volume can be explained by an overestimation of \intra\ interactions. The missing \inter\ interaction for non-vdW calculations is also reflected in the bulk modulus $B_0^\text{non-vdW}=3.86$\,GPa, which is approximately an order of magnitude smaller than the vdW values of $B_0^\text{\Grimme}=30.12$\,GPa and $B_0^\text{\TS}=33.12$\,GPa. The later are in the same region as the experimental values \cite{Millican.2009,Margadonna.2009}.

A minor effect is the volume expansion due to temperature. Since the experimental values are measured at room temperature, the volume at $T=0$\,K will be reduced. Indeed other works on 11-compounds show that the volume decreases with decreasing temperature \cite{Koz.2013}.

\subsection{Structural Properties}
Comparing the lattice parameters for non-vdW, the \Grimme\ and \TS\ corrections in Fig.~\ref{Fig:Structural_param_all} yield a significant improvement. Especially \zSeC\ is in nearly perfect agreement with the experiment value for both vdW implementations (see Fig.~\ref{Fig:Structural_param_all}~(e)). The origin is a reduction of the lattice parameter in \cd\ and the \ca\ ratio (see Fig.~\ref{Fig:Structural_param_all}~(b), (d)).
This reduction is a direct consequence of the improved description of the  \inter\ attraction by taking the vdW interaction correctly into account.

However, the lattice parameters of the vdW corrected calculations in \ad\ are $\approx 3\%$ smaller than the experimental ones. This has also consequences for the \ca\ ratio. Compared to non-vdW calculations it is reduced, due to the \inter\ layer effect of vdW interactions. Since at the same time the \intra\ interactins are overestimated, the \ca\ ratio is still larger than the experimental value. 

\begin{figure}[h]
	\centering
	\includegraphics[width=7cm]{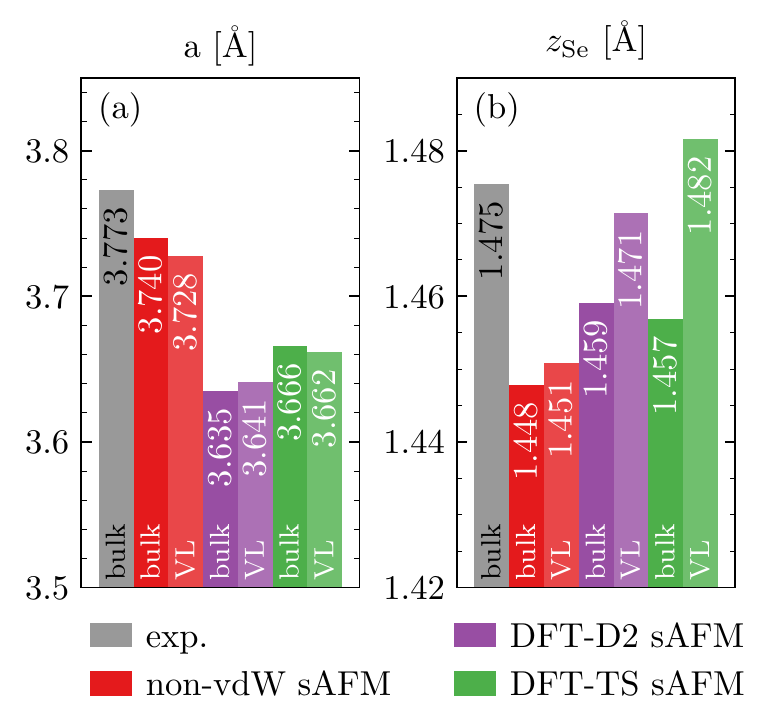}
	\caption{Lattice parameters for bulk FeSe and a FeSe vacuum layer (VL) in the sAFM state with and without vdW corrections. The experimental values by \cite{McQueen.2009} (gray), the numerical values for non-vdW sAFM calculations (red), sAFM \Grimme\ calculations (violet) and sAFM \TS\ calculations (green) are illustrated for (a) the lattice parameter in \ad\ and (b) the height of the Se-atom \zSe.}
	\label{Fig:Structural_param_all_1lay}
\end{figure}

To investigate the \intra\ interactions in detail, we additionally calculate an isolated FeSe layer in vacuum, i.\,e., a case where by construction no  \inter\ interactions occur. In Fig.~\ref{Fig:Structural_param_all_1lay}~(a) it can be seen, that the relaxed lattice parameter $a$ for the bulk compound and the vacuum layer (VL) do not show a significant difference. Thus the overbinding in \ad\ is mostly caused by the \intra\ vdW interactions. 
This result clearly shows that both approaches to include vdW bonding work well when other bond types such as covalent or ionic are absent, but do no consider the reduction of vdW interactions when they are present, as it is the case for \intra\ interactions.

While the intralayer lattice constant $a$ is almost identical for the bulk and the isolated layer, the two systems show a pronounced difference on the internal structure parameter \zSe. Switching the interlayer interaction off, systematicall increases this parameter compared to the bulk compound (see Fig.~\ref{Fig:Structural_param_all_1lay}~(b)). Thus, the \inter\ interactions cause a compression in \cd\ of the FeSe layers. Since these interactions are mostly driven by the vdW attraction of the Se-atoms, we consider the ratio \zSeC\ rather than the bare \zSe\ displacement to analyze structural improvement.

The investigation of the isolate layer shows that the lattice parameters in \ad\ (mainly driven by \intra\ non-vdW interactions) and \cd\ (mainly driven by \inter\ vdW interactions) are decoupled. Thus the non-vdW calculations provide a sufficiently accurate description of lattice parameter $a$, whereas  vdW corrected calculations are mandatory to describe the lattice parameter $c$. To simultaneously improve both quantities a vdW implementation would be needed that takes the effect of strong covalent bonds on the vdW interaction into account.
A possible approach to remove overbinding by VdW corrections is to perform the sum in Eq.~(\ref{eq:E_vdW}) only over the pairs of atoms that are in different layers. Since this requires a new implementation of vdW in the DFT code, we did not test it. 

\begin{figure*}[t]
	\centering
	\includegraphics[width=17.2cm]{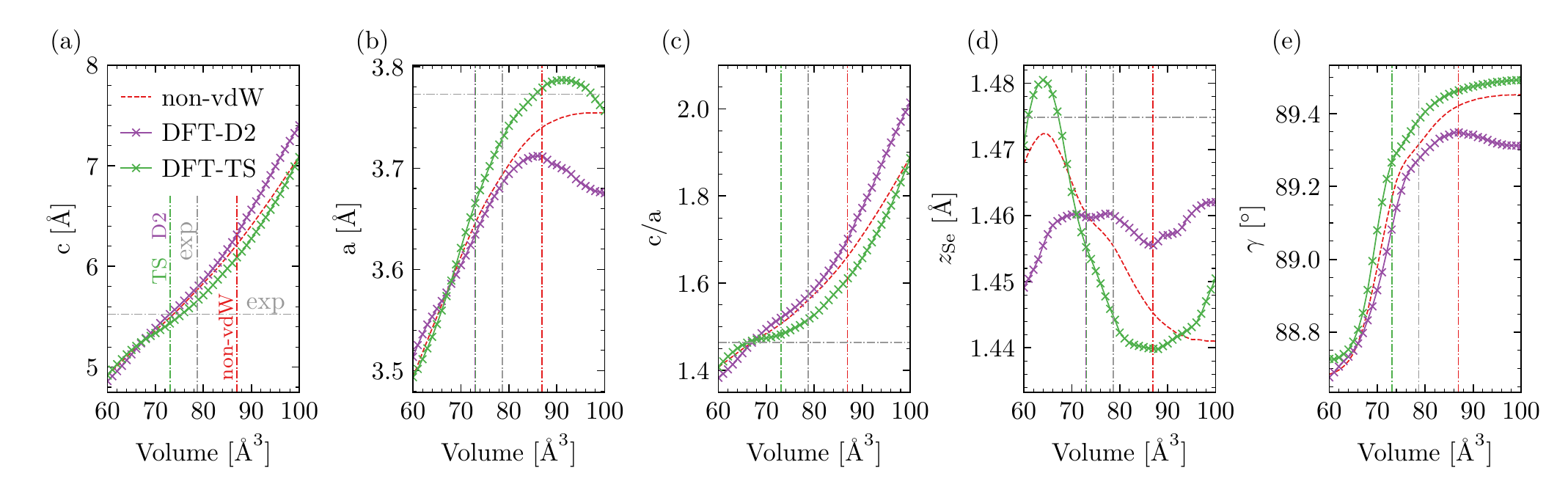}
	\caption{Volume dependence of the sAFM state for standard DFT calculations (red dashed) compared to vdW corrected ones by the \Grimme\ (violet solid) and the \TS\ (green solid) method of (a) the lattice parameter in \cd, (b) the lattice parameter in \ad, (c) the fraction of lattice parameter $c$ and $a$, (d) the height of Se-atoms above the iron-plane (\zSe) and (e) the angle between \alat\ and $b$. The dashed dotted lines indicate the experimental values by \cite{McQueen.2009} (grey), the non-vdW optimized volume (red) and the optimized ones for the \Grimme\ approach (violet) and the \TS\ approach (green).}
	\label{Fig:AFM_all_vdW}
\end{figure*}

Next we systematically analyze the structural impact of the volume reduction from the non-vdW optimized volume at $V^\text{non-vdW}_\text{min}=86.931$\AAA\  to the corrected volume (compare Fig.~\ref{Fig:AFM_energy_vdW}). As can be seen  in  Fig.~\ref{Fig:AFM_all_vdW} the difference between \Grimme\ and \TS\ is along this path more pronounced than the equilibrated values in Fig.~\ref{Fig:Structural_param_all}.

In the first place, the reduction in volume is connected with a reduction in the \ca\ ratio  to match the experimental values. As can be seen in Fig.~\ref{Fig:AFM_all_vdW}~(c), the \TS\ approach has a qualitatively different impact on the \ca\ ratio than the \Grimme\ approach. 
In the latter case it is enhanced next to $V^\text{non-vdW}$, what is caused by the treatment of the \intra\ interactions. This causes a reduction of the lattice parameter in \ad\ in Fig.~\ref{Fig:AFM_all_vdW}~(b) and consequently an increase of that one in \cd\ in Fig.~\ref{Fig:AFM_all_vdW}~(a). 
The overestimates of the \intra\ interaction is for this volume region less pronounced in the \TS\ approach. The increase of the lattice parameter $a$ might be caused by a Poisson effect due to the decreased $c$ value. The \zSe\ parameter in Fig.~\ref{Fig:AFM_all_vdW}~(d) shows a roughly anti-proportional trend compared to the lattice parameter $a$, thus the bond length tries to remain constant. A similar effect is also visible for the VL in Fig.~\ref{Fig:Structural_param_all_1lay}.

For even larger volumes, the lattice parameter $a$ and \zSe\ are nearly constant for the non-vdW curve, what is caused by vanishing \inter\ bonds. As the lattice parameter in \cd\ steadily increases for increasing volume, the  Fe-layers drift apart. This behavior corresponds to the well known exfoliation of graphene \cite{Allen.2010}, where vdW interactions also play a leading role. As $V^\text{non-vdW}_\text{min}$ is close to that region, it is most likely that the optimized lattice parameter of this approach are heavily influenced by those missing interactions.

When approaching the experimental volume at $V^\text{exp}_\text{min}=78.679$\AAA\ the slopes of the \ca\ ratios for the vdW approaches are reduced compared to the region of higher volumes. This effect is strongest for the \TS\ method. As the cell is compressed, the overestimation of the \intra\ vdW attraction causes now a stronger decrease of the lattice parameter $a$, whereas the value in \cd\ is less influenced. The same effect is also reflected in the steep slope of \zSe\ in Fig.~\ref{Fig:AFM_all_vdW}~(d). Here, \zSe\ for the \TS\ method varies a lot, while the value for the \Grimme\ approach does not show any significant changes. Moreover, for the \Grimme\ approach the values of the lattice parameter $a$ and the \ca\ ratio are in this volume region nearly the same as for the non-vdW approach. 

In the volume region around $V\approx70$\AAA, which  is close to the vdW optimized volumes of \Grimme\ and \TS, the structural parameters for the \Grimme\ and \TS\ method and in particular the \ca\ ratio are similar. Therefore, the effects of \inter\ and \intra\ vdW corrections are comparable for both approaches. The difference in the total energy, however, is still large (see. Fig.~\ref{Fig:AFM_energy_vdW}). 
For even smaller volumes the proximity of the second FeSe-layer causes several structural changes: At $V<65$\AAA\, the displacement \zSe\ decreases  for all approaches.
Note that  the overestimated \intra\ interactions cause also a squeezing of the Se-atoms in \cd\ as the attraction of \intra\ Se-up-atoms and Se-down-atoms is also overestimated. 
The kink visible for all approaches around $V=75$\AAA\  is caused by a tetragonal in-plane distortion ($\gamma\leq 89.4^\circ$, see Fig.~\ref{Fig:AFM_all_vdW}~(e)) seen for most AFM like structures in FeSe.

By looking at the complete volume region, the \Grimme\ and \TS\ approaches, i.e.,  their different treatment of the Hirschfeld partitioning and geometrical composition, imply significantly different behaviors. While the lattice parameters for the \Grimme\ agree for some volumes even less with experiment than the non-vdW ones, the values of the \TS\ approach are overall closer to the experiment. The \zSe\ parameter is within \Grimme\ for large volumes close to the experimental value, but the largely reduced gradient of \zSe\ with respect to volume as compared to \TS\ indicates a cancellation of interactions. 
Similar to the assessment of exchange correlation functionals in DFT \cite{Grabowski2007}, the good agreement of the vdW approaches close to their equilibrium volumes increases confidence in the vdW corrections. One should, however, keep in mind that the \intra\ interactions are  overestimated in both approaches.

\subsection{Paramagnetic van der Waals Calculations}
The PM vdW corrected calculations are similar to those from Sec.~\ref{Sec:PM}. In contrast to the non-vdW calculations, the minimization of the total energy is resulting into a well defined equilibrium volume, due to the correct description of the \inter\ binding. The converged optimal structural parameters are illustrated in Fig.~\ref{Fig:Structural_param_all}. The lattice parameters are comparable to those of the sAFM state. The \ca\ ratio is slightly enlarged, what can be related to the reduced \inter\ magnetic interactions already mentioned when discussing the non-vdW PM state. By looking at Fig.~\ref{Fig:Structural_param_all}~(e), it can be seen, that \zSeC\ does not show any significant changes compared to the sAFM state. This underlines again the similarity of both magnetic configurations. It also shows that the \inter\ attraction is solely caused by the presence of vdW interactions. It indicates that the \zSe\ coordinate does not depend on the magnetic structure and can only be correctly reproduced by DFT calculations with vdW corrections.

\subsection{Band Dispersion}
\begin{figure}[h]
	\centering
	\includegraphics[width=8.6cm]{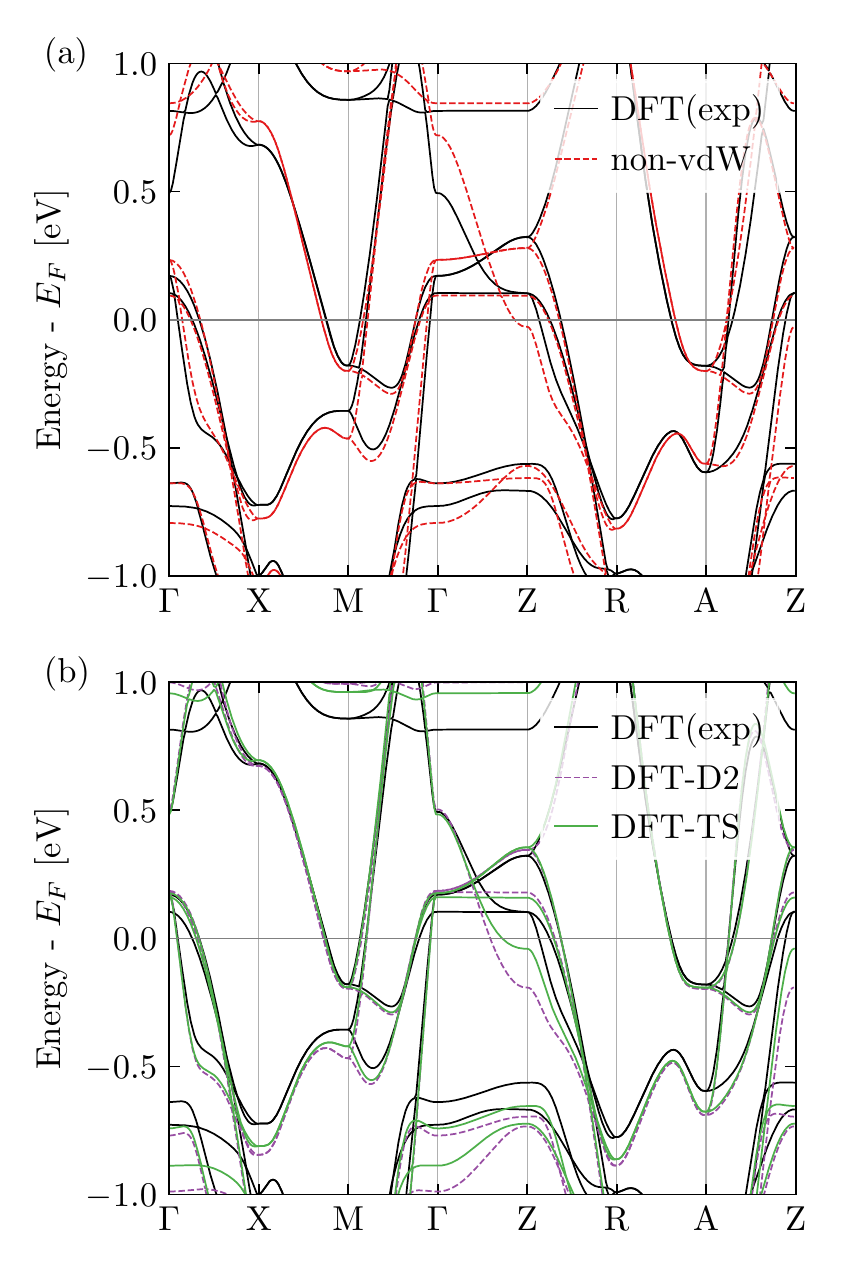}
	\caption{Electronic band dispersion near the Fermi energy obtained by non-magnetic DFT calculations for the different lattice parameters given in Fig.~\ref{Fig:Structural_param_all}. Specifically, non-vdW DFT calculation using experimental lattice parameters by \cite{McQueen.2009} (black solid) and (a) non-vdW sAFM state (red dashed) as well as (b) \Grimme\ vdW (violet dashed) and \TS\ vdW (green solid) corrections have bee used.}
	\label{Fig:BS_all}
\end{figure}

To investigate the electronic band dispersion we take the NM calculations with structural parameters obtained by experiments \cite{McQueen.2009} (see Fig.~\ref{Fig:Structural_param_all}) as a reference. Although, the NM state and the sAFM state differ for the structural parameters, their band dispersion is very similar except for some slight differences (see Fig.~\ref{Fig:BS_all}~(a)). For example, the $k_z$-dependence of Se $p_z$-orbital, which crosses the Fermi surface (FS) along $\Gamma \rightarrow Z$, is stronger for the experimental values, and the Fe $3d$-orbitals are mostly $k_z$ independent. As a result we find a similar shape of electron pockets around the M-point and the A-point of the Brillouin Zone and flatter dispersions from the $\Gamma$-point to the Z-point, what is due to $d_{xz}$- and $d_{yz}$-orbitals along $\Gamma \rightarrow Z$ direction.

For the vdW corrected calculations the changes in the real space structure are also reflected in the electronic dispersion. Comparing the band dispersion with and without vdW corrections, the sizable $k_z$-dependence of the Fe $3d$-orbitals is indeed found, consistent with previous works \cite{Ricci.2013}. Additionally, the shape of the electron pockets for $k_z=\pi$ is the same as for calculations with the experimental lattice parameters, a clear improvement to non-vdW sAFM calculations. Although other works propose \inter\ driven effects on the band dispersion to be less important for superconductivity, a reconstruction of the FS by the Se $p_z$-orbital dominated band from $\Gamma\rightarrow Z$ is found \cite{Guterding.2017b}. This may be related to an overestimation of \inter\ Se-attraction.

Comparing both vdW approaches for FeSe (presumptive for all 11-based FeSC), the electronic band dispersion turns out to be very similar. Most important is that the enhancement of the \inter\ interaction, which is in agreement with considerations of other works claiming vdW interactions to be important for FS reconstructions \cite{Guterding.2017b}. 
The \TS\ approach includes the impact of the local environment on the vdW interaction, slightly reducing the overestimation of the \intra\ interaction in \Grimme\ and improving the \zSeC\ ratio. The only noticeable consequence in the band dispersion, however, is a modified band energy near the FS at the Z point. 

\section{Summary} \label{Sec:Sum}

Based on a systematic DFT study, we show that the delicate interplay between the real space structure of FeSe and the resulting electronic dispersion requires the inclusion of vdW interactions and spin-disorder (PM). The introduced \inter\ attraction makes FeSe a much more 3-dimensional material in \ai\ calculations and agrees with recent ARPES experiments \cite{Watson.2015}. For the ground state the two most common vdW approaches \Grimme\ and \TS\ yield similar lattice parameters at ambient conditions, yet for the pressure-dependent calculations \TS\ is more appropriate, as it takes the real space local environment to estimate the vdW strength. Those investigations underline the fact that charge driven Se-Se interactions play an important role, similar to the the magnetic Fe-Fe interactions. The calculated lattice parameters using vdW and PM effects show the need for \inter\ vdW corrections. For both magnetic configurations, sAFM and PM, the height of the Se-atoms \zSeC\ are mostly identical, substantiating a lack of \inter\ attraction to be responsible for the lattice mismatch of all previous \ai\ approaches. We also show that the \intra\ coupling do not include vdW interactions, however, the presence of them is negligible for the electronic dispersion. To further improve the lattice parameters, a vdW correction neglecting all \intra\ contributions is required, as we showed that the mismatch in \ad\ is mostly driven by overestimated \intra\ attraction. We also conclude, that the used magnetic structure will be less important for the lattice parameters, if the overall magnetic moment is close to zero. It indicates that the \zSe\ coordinate is actually independent of the magnetic structure and can only be correctly reproduced by DFT calculations with vdW corrections. Moreover, the fact that those magnetic structures lead to the very same structural properties substantiate the presence of competing magnetic orders in the nematic phase \cite{Martiny.2019, Liu.2016}. Since those structural properties are essential for the superconductivity in FeSC, the improvement of the \ai\ methods obtained in this work can be important for its understanding. The change of the character of the electronic dispersion from quasi two-dimensional towards more three-dimensional one with a significant contribution of the Se $p_z$-orbitals is also relevant in the context of topological features, discussed previously \cite{Zhang.2018d} and needs to be explored further both experimentally and theoretically.


\bibliographystyle{apsrev4-1}
\bibliography{bib_FeSe_vdW}

\begin{thebibliography}{56}%
\makeatletter
\providecommand \@ifxundefined [1]{%
 \@ifx{#1\undefined}
}%
\providecommand \@ifnum [1]{%
 \ifnum #1\expandafter \@firstoftwo
 \else \expandafter \@secondoftwo
 \fi
}%
\providecommand \@ifx [1]{%
 \ifx #1\expandafter \@firstoftwo
 \else \expandafter \@secondoftwo
 \fi
}%
\providecommand \natexlab [1]{#1}%
\providecommand \enquote  [1]{``#1''}%
\providecommand \bibnamefont  [1]{#1}%
\providecommand \bibfnamefont [1]{#1}%
\providecommand \citenamefont [1]{#1}%
\providecommand \href@noop [0]{\@secondoftwo}%
\providecommand \href [0]{\begingroup \@sanitize@url \@href}%
\providecommand \@href[1]{\@@startlink{#1}\@@href}%
\providecommand \@@href[1]{\endgroup#1\@@endlink}%
\providecommand \@sanitize@url [0]{\catcode `\\12\catcode `\$12\catcode
  `\&12\catcode `\#12\catcode `\^12\catcode `\_12\catcode `\%12\relax}%
\providecommand \@@startlink[1]{}%
\providecommand \@@endlink[0]{}%
\providecommand \url  [0]{\begingroup\@sanitize@url \@url }%
\providecommand \@url [1]{\endgroup\@href {#1}{\urlprefix }}%
\providecommand \urlprefix  [0]{URL }%
\providecommand \Eprint [0]{\href }%
\providecommand \doibase [0]{http://dx.doi.org/}%
\providecommand \selectlanguage [0]{\@gobble}%
\providecommand \bibinfo  [0]{\@secondoftwo}%
\providecommand \bibfield  [0]{\@secondoftwo}%
\providecommand \translation [1]{[#1]}%
\providecommand \BibitemOpen [0]{}%
\providecommand \bibitemStop [0]{}%
\providecommand \bibitemNoStop [0]{.\EOS\space}%
\providecommand \EOS [0]{\spacefactor3000\relax}%
\providecommand \BibitemShut  [1]{\csname bibitem#1\endcsname}%
\let\auto@bib@innerbib\@empty
\bibitem [{\citenamefont {B{\"o}hmer}\ and\ \citenamefont
  {Kreisel}(2018)}]{Bohmer.2018}%
  \BibitemOpen
  \bibfield  {author} {\bibinfo {author} {\bibfnamefont {A.~E.}\ \bibnamefont
  {B{\"o}hmer}}\ and\ \bibinfo {author} {\bibfnamefont {A.}~\bibnamefont
  {Kreisel}},\ }\href {\doibase 10.1088/1361-648X/aa9caa} {\bibfield  {journal}
  {\bibinfo  {journal} {Journal of physics. Condensed matter : an Institute of
  Physics journal}\ }\textbf {\bibinfo {volume} {30}},\ \bibinfo {pages}
  {023001} (\bibinfo {year} {2018})}\BibitemShut {NoStop}%
\bibitem [{\citenamefont {McQueen}\ \emph
  {et~al.}(2009{\natexlab{a}})\citenamefont {McQueen}, \citenamefont
  {Williams}, \citenamefont {Stephens}, \citenamefont {Tao}, \citenamefont
  {Zhu}, \citenamefont {Ksenofontov}, \citenamefont {Casper}, \citenamefont
  {Felser},\ and\ \citenamefont {Cava}}]{McQueen.2009b}%
  \BibitemOpen
  \bibfield  {author} {\bibinfo {author} {\bibfnamefont {T.~M.}\ \bibnamefont
  {McQueen}}, \bibinfo {author} {\bibfnamefont {A.~J.}\ \bibnamefont
  {Williams}}, \bibinfo {author} {\bibfnamefont {P.~W.}\ \bibnamefont
  {Stephens}}, \bibinfo {author} {\bibfnamefont {J.}~\bibnamefont {Tao}},
  \bibinfo {author} {\bibfnamefont {Y.}~\bibnamefont {Zhu}}, \bibinfo {author}
  {\bibfnamefont {V.}~\bibnamefont {Ksenofontov}}, \bibinfo {author}
  {\bibfnamefont {F.}~\bibnamefont {Casper}}, \bibinfo {author} {\bibfnamefont
  {C.}~\bibnamefont {Felser}}, \ and\ \bibinfo {author} {\bibfnamefont {R.~J.}\
  \bibnamefont {Cava}},\ }\href {\doibase 10.1103/PhysRevLett.103.057002}
  {\bibfield  {journal} {\bibinfo  {journal} {Physical Review Letters}\
  }\textbf {\bibinfo {volume} {103}},\ \bibinfo {pages} {057002} (\bibinfo
  {year} {2009}{\natexlab{a}})}\BibitemShut {NoStop}%
\bibitem [{\citenamefont {Margadonna}\ \emph {et~al.}(2008)\citenamefont
  {Margadonna}, \citenamefont {Takabayashi}, \citenamefont {McDonald},
  \citenamefont {Kasperkiewicz}, \citenamefont {Mizuguchi}, \citenamefont
  {Takano}, \citenamefont {Fitch}, \citenamefont {Suard},\ and\ \citenamefont
  {Prassides}}]{Margadonna.2008}%
  \BibitemOpen
  \bibfield  {author} {\bibinfo {author} {\bibfnamefont {S.}~\bibnamefont
  {Margadonna}}, \bibinfo {author} {\bibfnamefont {Y.}~\bibnamefont
  {Takabayashi}}, \bibinfo {author} {\bibfnamefont {M.~T.}\ \bibnamefont
  {McDonald}}, \bibinfo {author} {\bibfnamefont {K.}~\bibnamefont
  {Kasperkiewicz}}, \bibinfo {author} {\bibfnamefont {Y.}~\bibnamefont
  {Mizuguchi}}, \bibinfo {author} {\bibfnamefont {Y.}~\bibnamefont {Takano}},
  \bibinfo {author} {\bibfnamefont {A.~N.}\ \bibnamefont {Fitch}}, \bibinfo
  {author} {\bibfnamefont {E.}~\bibnamefont {Suard}}, \ and\ \bibinfo {author}
  {\bibfnamefont {K.}~\bibnamefont {Prassides}},\ }\href {\doibase
  10.1039/b813076k} {\bibfield  {journal} {\bibinfo  {journal} {Chemical
  Communications (Cambridge, England)}\ }\textbf {\bibinfo {volume} {0}},\
  \bibinfo {pages} {5607} (\bibinfo {year} {2008})}\BibitemShut {NoStop}%
\bibitem [{\citenamefont {Fernandes}\ and\ \citenamefont
  {Schmalian}(2012)}]{Fernandes.2012b}%
  \BibitemOpen
  \bibfield  {author} {\bibinfo {author} {\bibfnamefont {R.~M.}\ \bibnamefont
  {Fernandes}}\ and\ \bibinfo {author} {\bibfnamefont {J.}~\bibnamefont
  {Schmalian}},\ }\href {\doibase 10.1088/0953-2048/25/8/084005} {\bibfield
  {journal} {\bibinfo  {journal} {Superconductor Science and Technology}\
  }\textbf {\bibinfo {volume} {25}},\ \bibinfo {pages} {084005} (\bibinfo
  {year} {2012})}\BibitemShut {NoStop}%
\bibitem [{\citenamefont {Bendele}\ \emph {et~al.}(2010)\citenamefont
  {Bendele}, \citenamefont {Amato}, \citenamefont {Conder}, \citenamefont
  {Elender}, \citenamefont {Keller}, \citenamefont {Klauss}, \citenamefont
  {Luetkens}, \citenamefont {Pomjakushina}, \citenamefont {Raselli},\ and\
  \citenamefont {Khasanov}}]{Bendele.2010}%
  \BibitemOpen
  \bibfield  {author} {\bibinfo {author} {\bibfnamefont {M.}~\bibnamefont
  {Bendele}}, \bibinfo {author} {\bibfnamefont {A.}~\bibnamefont {Amato}},
  \bibinfo {author} {\bibfnamefont {K.}~\bibnamefont {Conder}}, \bibinfo
  {author} {\bibfnamefont {M.}~\bibnamefont {Elender}}, \bibinfo {author}
  {\bibfnamefont {H.}~\bibnamefont {Keller}}, \bibinfo {author} {\bibfnamefont
  {H.-H.}\ \bibnamefont {Klauss}}, \bibinfo {author} {\bibfnamefont
  {H.}~\bibnamefont {Luetkens}}, \bibinfo {author} {\bibfnamefont
  {E.}~\bibnamefont {Pomjakushina}}, \bibinfo {author} {\bibfnamefont
  {A.}~\bibnamefont {Raselli}}, \ and\ \bibinfo {author} {\bibfnamefont
  {R.}~\bibnamefont {Khasanov}},\ }\href {\doibase
  10.1103/PhysRevLett.104.087003} {\bibfield  {journal} {\bibinfo  {journal}
  {Physical Review Letters}\ }\textbf {\bibinfo {volume} {104}},\ \bibinfo
  {pages} {087003} (\bibinfo {year} {2010})}\BibitemShut {NoStop}%
\bibitem [{\citenamefont {Medvedev}\ \emph {et~al.}(2009)\citenamefont
  {Medvedev}, \citenamefont {McQueen}, \citenamefont {Troyan}, \citenamefont
  {Palasyuk}, \citenamefont {Eremets}, \citenamefont {Cava}, \citenamefont
  {Naghavi}, \citenamefont {Casper}, \citenamefont {Ksenofontov}, \citenamefont
  {Wortmann},\ and\ \citenamefont {Felser}}]{Medvedev.2009}%
  \BibitemOpen
  \bibfield  {author} {\bibinfo {author} {\bibfnamefont {S.}~\bibnamefont
  {Medvedev}}, \bibinfo {author} {\bibfnamefont {T.~M.}\ \bibnamefont
  {McQueen}}, \bibinfo {author} {\bibfnamefont {I.~A.}\ \bibnamefont {Troyan}},
  \bibinfo {author} {\bibfnamefont {T.}~\bibnamefont {Palasyuk}}, \bibinfo
  {author} {\bibfnamefont {M.~I.}\ \bibnamefont {Eremets}}, \bibinfo {author}
  {\bibfnamefont {R.~J.}\ \bibnamefont {Cava}}, \bibinfo {author}
  {\bibfnamefont {S.}~\bibnamefont {Naghavi}}, \bibinfo {author} {\bibfnamefont
  {F.}~\bibnamefont {Casper}}, \bibinfo {author} {\bibfnamefont
  {V.}~\bibnamefont {Ksenofontov}}, \bibinfo {author} {\bibfnamefont
  {G.}~\bibnamefont {Wortmann}}, \ and\ \bibinfo {author} {\bibfnamefont
  {C.}~\bibnamefont {Felser}},\ }\href {\doibase 10.1038/NMAT2491} {\bibfield
  {journal} {\bibinfo  {journal} {Nature Materials}\ }\textbf {\bibinfo
  {volume} {8}},\ \bibinfo {pages} {630} (\bibinfo {year} {2009})}\BibitemShut
  {NoStop}%
\bibitem [{\citenamefont {Bendele}\ \emph {et~al.}(2012)\citenamefont
  {Bendele}, \citenamefont {Ichsanow}, \citenamefont {Pashkevich},
  \citenamefont {Keller}, \citenamefont {Str{\"a}ssle}, \citenamefont {Gusev},
  \citenamefont {Pomjakushina}, \citenamefont {Conder}, \citenamefont
  {Khasanov},\ and\ \citenamefont {Keller}}]{Bendele.2012}%
  \BibitemOpen
  \bibfield  {author} {\bibinfo {author} {\bibfnamefont {M.}~\bibnamefont
  {Bendele}}, \bibinfo {author} {\bibfnamefont {A.}~\bibnamefont {Ichsanow}},
  \bibinfo {author} {\bibfnamefont {Y.}~\bibnamefont {Pashkevich}}, \bibinfo
  {author} {\bibfnamefont {L.}~\bibnamefont {Keller}}, \bibinfo {author}
  {\bibfnamefont {T.}~\bibnamefont {Str{\"a}ssle}}, \bibinfo {author}
  {\bibfnamefont {A.}~\bibnamefont {Gusev}}, \bibinfo {author} {\bibfnamefont
  {E.}~\bibnamefont {Pomjakushina}}, \bibinfo {author} {\bibfnamefont
  {K.}~\bibnamefont {Conder}}, \bibinfo {author} {\bibfnamefont
  {R.}~\bibnamefont {Khasanov}}, \ and\ \bibinfo {author} {\bibfnamefont
  {H.}~\bibnamefont {Keller}},\ }\href {\doibase 10.1103/PhysRevB.85.064517}
  {\bibfield  {journal} {\bibinfo  {journal} {Physical Review B}\ }\textbf
  {\bibinfo {volume} {85}},\ \bibinfo {pages} {064517} (\bibinfo {year}
  {2012})}\BibitemShut {NoStop}%
\bibitem [{\citenamefont {Sun}\ \emph {et~al.}(2016)\citenamefont {Sun},
  \citenamefont {Matsuura}, \citenamefont {Ye}, \citenamefont {Mizukami},
  \citenamefont {Shimozawa}, \citenamefont {Matsubayashi}, \citenamefont
  {Yamashita}, \citenamefont {Watashige}, \citenamefont {Kasahara},
  \citenamefont {Matsuda}, \citenamefont {Yan}, \citenamefont {Sales},
  \citenamefont {Uwatoko}, \citenamefont {Cheng},\ and\ \citenamefont
  {Shibauchi}}]{Sun.2016}%
  \BibitemOpen
  \bibfield  {author} {\bibinfo {author} {\bibfnamefont {J.~P.}\ \bibnamefont
  {Sun}}, \bibinfo {author} {\bibfnamefont {K.}~\bibnamefont {Matsuura}},
  \bibinfo {author} {\bibfnamefont {G.~Z.}\ \bibnamefont {Ye}}, \bibinfo
  {author} {\bibfnamefont {Y.}~\bibnamefont {Mizukami}}, \bibinfo {author}
  {\bibfnamefont {M.}~\bibnamefont {Shimozawa}}, \bibinfo {author}
  {\bibfnamefont {K.}~\bibnamefont {Matsubayashi}}, \bibinfo {author}
  {\bibfnamefont {M.}~\bibnamefont {Yamashita}}, \bibinfo {author}
  {\bibfnamefont {T.}~\bibnamefont {Watashige}}, \bibinfo {author}
  {\bibfnamefont {S.}~\bibnamefont {Kasahara}}, \bibinfo {author}
  {\bibfnamefont {Y.}~\bibnamefont {Matsuda}}, \bibinfo {author} {\bibfnamefont
  {J.-Q.}\ \bibnamefont {Yan}}, \bibinfo {author} {\bibfnamefont {B.~C.}\
  \bibnamefont {Sales}}, \bibinfo {author} {\bibfnamefont {Y.}~\bibnamefont
  {Uwatoko}}, \bibinfo {author} {\bibfnamefont {J.-G.}\ \bibnamefont {Cheng}},
  \ and\ \bibinfo {author} {\bibfnamefont {T.}~\bibnamefont {Shibauchi}},\
  }\href {\doibase 10.1038/ncomms12146} {\bibfield  {journal} {\bibinfo
  {journal} {Nature Communications}\ }\textbf {\bibinfo {volume} {7}},\
  \bibinfo {pages} {12146} (\bibinfo {year} {2016})}\BibitemShut {NoStop}%
\bibitem [{\citenamefont {Kothapalli}\ \emph {et~al.}(2016)\citenamefont
  {Kothapalli}, \citenamefont {B{\"o}hmer}, \citenamefont {Jayasekara},
  \citenamefont {Ueland}, \citenamefont {Das}, \citenamefont {Sapkota},
  \citenamefont {Taufour}, \citenamefont {Xiao}, \citenamefont {Alp},
  \citenamefont {Bud'ko}, \citenamefont {Canfield}, \citenamefont {Kreyssig},\
  and\ \citenamefont {Goldman}}]{Kothapalli.2016}%
  \BibitemOpen
  \bibfield  {author} {\bibinfo {author} {\bibfnamefont {K.}~\bibnamefont
  {Kothapalli}}, \bibinfo {author} {\bibfnamefont {A.~E.}\ \bibnamefont
  {B{\"o}hmer}}, \bibinfo {author} {\bibfnamefont {W.~T.}\ \bibnamefont
  {Jayasekara}}, \bibinfo {author} {\bibfnamefont {B.~G.}\ \bibnamefont
  {Ueland}}, \bibinfo {author} {\bibfnamefont {P.}~\bibnamefont {Das}},
  \bibinfo {author} {\bibfnamefont {A.}~\bibnamefont {Sapkota}}, \bibinfo
  {author} {\bibfnamefont {V.}~\bibnamefont {Taufour}}, \bibinfo {author}
  {\bibfnamefont {Y.}~\bibnamefont {Xiao}}, \bibinfo {author} {\bibfnamefont
  {E.}~\bibnamefont {Alp}}, \bibinfo {author} {\bibfnamefont {S.~L.}\
  \bibnamefont {Bud'ko}}, \bibinfo {author} {\bibfnamefont {P.~C.}\
  \bibnamefont {Canfield}}, \bibinfo {author} {\bibfnamefont {A.}~\bibnamefont
  {Kreyssig}}, \ and\ \bibinfo {author} {\bibfnamefont {A.~I.}\ \bibnamefont
  {Goldman}},\ }\href {\doibase 10.1038/ncomms12728} {\bibfield  {journal}
  {\bibinfo  {journal} {Nature Communications}\ }\textbf {\bibinfo {volume}
  {7}},\ \bibinfo {pages} {12728} (\bibinfo {year} {2016})}\BibitemShut
  {NoStop}%
\bibitem [{\citenamefont {Guterding}\ \emph
  {et~al.}(2017{\natexlab{a}})\citenamefont {Guterding}, \citenamefont
  {Backes}, \citenamefont {Tomi{\'c}}, \citenamefont {Jeschke},\ and\
  \citenamefont {Valent{\'i}}}]{Guterding.2017}%
  \BibitemOpen
  \bibfield  {author} {\bibinfo {author} {\bibfnamefont {D.}~\bibnamefont
  {Guterding}}, \bibinfo {author} {\bibfnamefont {S.}~\bibnamefont {Backes}},
  \bibinfo {author} {\bibfnamefont {M.}~\bibnamefont {Tomi{\'c}}}, \bibinfo
  {author} {\bibfnamefont {H.~O.}\ \bibnamefont {Jeschke}}, \ and\ \bibinfo
  {author} {\bibfnamefont {R.}~\bibnamefont {Valent{\'i}}},\ }\href {\doibase
  10.1002/pssb.201600164} {\bibfield  {journal} {\bibinfo  {journal} {physica
  status solidi (b)}\ }\textbf {\bibinfo {volume} {254}},\ \bibinfo {pages}
  {1600164} (\bibinfo {year} {2017}{\natexlab{a}})}\BibitemShut {NoStop}%
\bibitem [{\citenamefont {Eschrig}\ and\ \citenamefont
  {Koepernik}(2009)}]{Eschrig.2009}%
  \BibitemOpen
  \bibfield  {author} {\bibinfo {author} {\bibfnamefont {H.}~\bibnamefont
  {Eschrig}}\ and\ \bibinfo {author} {\bibfnamefont {K.}~\bibnamefont
  {Koepernik}},\ }\href {\doibase 10.1103/PhysRevB.80.104503} {\bibfield
  {journal} {\bibinfo  {journal} {Physical Review B}\ }\textbf {\bibinfo
  {volume} {80}},\ \bibinfo {pages} {104503} (\bibinfo {year}
  {2009})}\BibitemShut {NoStop}%
\bibitem [{\citenamefont {Lochner}\ \emph {et~al.}(2017)\citenamefont
  {Lochner}, \citenamefont {Ahn}, \citenamefont {Hickel},\ and\ \citenamefont
  {Eremin}}]{Lochner.2017}%
  \BibitemOpen
  \bibfield  {author} {\bibinfo {author} {\bibfnamefont {F.}~\bibnamefont
  {Lochner}}, \bibinfo {author} {\bibfnamefont {F.}~\bibnamefont {Ahn}},
  \bibinfo {author} {\bibfnamefont {T.}~\bibnamefont {Hickel}}, \ and\ \bibinfo
  {author} {\bibfnamefont {I.}~\bibnamefont {Eremin}},\ }\href {\doibase
  10.1103/PhysRevB.96.094521} {\bibfield  {journal} {\bibinfo  {journal}
  {Physical Review B}\ }\textbf {\bibinfo {volume} {96}},\ \bibinfo {pages}
  {094521} (\bibinfo {year} {2017})}\BibitemShut {NoStop}%
\bibitem [{\citenamefont {Johnston}(2010)}]{Johnston.2010}%
  \BibitemOpen
  \bibfield  {author} {\bibinfo {author} {\bibfnamefont {D.~C.}\ \bibnamefont
  {Johnston}},\ }\href {\doibase 10.1080/00018732.2010.513480} {\bibfield
  {journal} {\bibinfo  {journal} {Advances in Physics}\ }\textbf {\bibinfo
  {volume} {59}},\ \bibinfo {pages} {803} (\bibinfo {year} {2010})}\BibitemShut
  {NoStop}%
\bibitem [{\citenamefont {Hosono}\ \emph {et~al.}(2018)\citenamefont {Hosono},
  \citenamefont {Yamamoto}, \citenamefont {Hiramatsu},\ and\ \citenamefont
  {Ma}}]{Hosono.2018}%
  \BibitemOpen
  \bibfield  {author} {\bibinfo {author} {\bibfnamefont {H.}~\bibnamefont
  {Hosono}}, \bibinfo {author} {\bibfnamefont {A.}~\bibnamefont {Yamamoto}},
  \bibinfo {author} {\bibfnamefont {H.}~\bibnamefont {Hiramatsu}}, \ and\
  \bibinfo {author} {\bibfnamefont {Y.}~\bibnamefont {Ma}},\ }\href {\doibase
  10.1016/j.mattod.2017.09.006} {\bibfield  {journal} {\bibinfo  {journal}
  {Materials Today}\ }\textbf {\bibinfo {volume} {21}},\ \bibinfo {pages} {278}
  (\bibinfo {year} {2018})}\BibitemShut {NoStop}%
\bibitem [{\citenamefont {Dai}(2015)}]{Dai.2015}%
  \BibitemOpen
  \bibfield  {author} {\bibinfo {author} {\bibfnamefont {P.}~\bibnamefont
  {Dai}},\ }\href {\doibase 10.1103/RevModPhys.87.855} {\bibfield  {journal}
  {\bibinfo  {journal} {Reviews of Modern Physics}\ }\textbf {\bibinfo {volume}
  {87}},\ \bibinfo {pages} {855} (\bibinfo {year} {2015})}\BibitemShut
  {NoStop}%
\bibitem [{\citenamefont {Li}\ \emph {et~al.}(2009)\citenamefont {Li},
  \citenamefont {Zhu}, \citenamefont {Guo}, \citenamefont {Xu},\ and\
  \citenamefont {Liu}}]{Li.2009b}%
  \BibitemOpen
  \bibfield  {author} {\bibinfo {author} {\bibfnamefont {Y.-F.}\ \bibnamefont
  {Li}}, \bibinfo {author} {\bibfnamefont {L.-F.}\ \bibnamefont {Zhu}},
  \bibinfo {author} {\bibfnamefont {S.-D.}\ \bibnamefont {Guo}}, \bibinfo
  {author} {\bibfnamefont {Y.-C.}\ \bibnamefont {Xu}}, \ and\ \bibinfo {author}
  {\bibfnamefont {B.-G.}\ \bibnamefont {Liu}},\ }\href {\doibase
  10.1088/0953-8984/21/11/115701} {\bibfield  {journal} {\bibinfo  {journal}
  {Journal of physics. Condensed matter : an Institute of Physics journal}\
  }\textbf {\bibinfo {volume} {21}},\ \bibinfo {pages} {115701} (\bibinfo
  {year} {2009})}\BibitemShut {NoStop}%
\bibitem [{\citenamefont {Khasanov}\ \emph {et~al.}(2008)\citenamefont
  {Khasanov}, \citenamefont {Conder}, \citenamefont {Pomjakushina},
  \citenamefont {Amato}, \citenamefont {Baines}, \citenamefont {Bukowski},
  \citenamefont {Karpinski}, \citenamefont {Katrych}, \citenamefont {Klauss},
  \citenamefont {Luetkens}, \citenamefont {Shengelaya},\ and\ \citenamefont
  {Zhigadlo}}]{Khasanov.2008}%
  \BibitemOpen
  \bibfield  {author} {\bibinfo {author} {\bibfnamefont {R.}~\bibnamefont
  {Khasanov}}, \bibinfo {author} {\bibfnamefont {K.}~\bibnamefont {Conder}},
  \bibinfo {author} {\bibfnamefont {E.}~\bibnamefont {Pomjakushina}}, \bibinfo
  {author} {\bibfnamefont {A.}~\bibnamefont {Amato}}, \bibinfo {author}
  {\bibfnamefont {C.}~\bibnamefont {Baines}}, \bibinfo {author} {\bibfnamefont
  {Z.}~\bibnamefont {Bukowski}}, \bibinfo {author} {\bibfnamefont
  {J.}~\bibnamefont {Karpinski}}, \bibinfo {author} {\bibfnamefont
  {S.}~\bibnamefont {Katrych}}, \bibinfo {author} {\bibfnamefont {H.-H.}\
  \bibnamefont {Klauss}}, \bibinfo {author} {\bibfnamefont {H.}~\bibnamefont
  {Luetkens}}, \bibinfo {author} {\bibfnamefont {A.}~\bibnamefont
  {Shengelaya}}, \ and\ \bibinfo {author} {\bibfnamefont {N.~D.}\ \bibnamefont
  {Zhigadlo}},\ }\href {\doibase 10.1103/PhysRevB.78.220510} {\bibfield
  {journal} {\bibinfo  {journal} {Physical Review B}\ }\textbf {\bibinfo
  {volume} {78}},\ \bibinfo {pages} {220510} (\bibinfo {year}
  {2008})}\BibitemShut {NoStop}%
\bibitem [{\citenamefont {Hsu}\ \emph {et~al.}(2008)\citenamefont {Hsu},
  \citenamefont {Luo}, \citenamefont {Yeh}, \citenamefont {Chen}, \citenamefont
  {Huang}, \citenamefont {Wu}, \citenamefont {Lee}, \citenamefont {Huang},
  \citenamefont {Chu}, \citenamefont {Yan},\ and\ \citenamefont
  {Wu}}]{Hsu.2008}%
  \BibitemOpen
  \bibfield  {author} {\bibinfo {author} {\bibfnamefont {F.-C.}\ \bibnamefont
  {Hsu}}, \bibinfo {author} {\bibfnamefont {J.-Y.}\ \bibnamefont {Luo}},
  \bibinfo {author} {\bibfnamefont {K.-W.}\ \bibnamefont {Yeh}}, \bibinfo
  {author} {\bibfnamefont {T.-K.}\ \bibnamefont {Chen}}, \bibinfo {author}
  {\bibfnamefont {T.-W.}\ \bibnamefont {Huang}}, \bibinfo {author}
  {\bibfnamefont {P.~M.}\ \bibnamefont {Wu}}, \bibinfo {author} {\bibfnamefont
  {Y.-C.}\ \bibnamefont {Lee}}, \bibinfo {author} {\bibfnamefont {Y.-L.}\
  \bibnamefont {Huang}}, \bibinfo {author} {\bibfnamefont {Y.-Y.}\ \bibnamefont
  {Chu}}, \bibinfo {author} {\bibfnamefont {D.-C.}\ \bibnamefont {Yan}}, \ and\
  \bibinfo {author} {\bibfnamefont {M.-K.}\ \bibnamefont {Wu}},\ }\href
  {\doibase 10.1073/pnas.0807325105} {\bibfield  {journal} {\bibinfo  {journal}
  {Proceedings of the National Academy of Sciences of the United States of
  America}\ }\textbf {\bibinfo {volume} {105}},\ \bibinfo {pages} {14262}
  (\bibinfo {year} {2008})}\BibitemShut {NoStop}%
\bibitem [{\citenamefont {Liu}\ \emph {et~al.}(2016)\citenamefont {Liu},
  \citenamefont {Lu},\ and\ \citenamefont {Xiang}}]{Liu.2016}%
  \BibitemOpen
  \bibfield  {author} {\bibinfo {author} {\bibfnamefont {K.}~\bibnamefont
  {Liu}}, \bibinfo {author} {\bibfnamefont {Z.-Y.}\ \bibnamefont {Lu}}, \ and\
  \bibinfo {author} {\bibfnamefont {T.}~\bibnamefont {Xiang}},\ }\href
  {\doibase 10.1103/PhysRevB.93.205154} {\bibfield  {journal} {\bibinfo
  {journal} {Physical Review B}\ }\textbf {\bibinfo {volume} {93}},\ \bibinfo
  {pages} {205154} (\bibinfo {year} {2016})}\BibitemShut {NoStop}%
\bibitem [{\citenamefont {Glasbrenner}\ \emph {et~al.}(2015)\citenamefont
  {Glasbrenner}, \citenamefont {Mazin}, \citenamefont {Jeschke}, \citenamefont
  {Hirschfeld}, \citenamefont {Fernandes},\ and\ \citenamefont
  {Valent{\'i}}}]{Glasbrenner.2015}%
  \BibitemOpen
  \bibfield  {author} {\bibinfo {author} {\bibfnamefont {J.~K.}\ \bibnamefont
  {Glasbrenner}}, \bibinfo {author} {\bibfnamefont {I.~I.}\ \bibnamefont
  {Mazin}}, \bibinfo {author} {\bibfnamefont {H.~O.}\ \bibnamefont {Jeschke}},
  \bibinfo {author} {\bibfnamefont {P.~J.}\ \bibnamefont {Hirschfeld}},
  \bibinfo {author} {\bibfnamefont {R.~M.}\ \bibnamefont {Fernandes}}, \ and\
  \bibinfo {author} {\bibfnamefont {R.}~\bibnamefont {Valent{\'i}}},\ }\href
  {\doibase 10.1038/nphys3434} {\bibfield  {journal} {\bibinfo  {journal}
  {Nature Physics}\ }\textbf {\bibinfo {volume} {11}},\ \bibinfo {pages} {953}
  (\bibinfo {year} {2015})}\BibitemShut {NoStop}%
\bibitem [{\citenamefont {Christensen}\ \emph {et~al.}(2019)\citenamefont
  {Christensen}, \citenamefont {Kang},\ and\ \citenamefont
  {Fernandes}}]{Christensen.2019}%
  \BibitemOpen
  \bibfield  {author} {\bibinfo {author} {\bibfnamefont {M.~H.}\ \bibnamefont
  {Christensen}}, \bibinfo {author} {\bibfnamefont {J.}~\bibnamefont {Kang}}, \
  and\ \bibinfo {author} {\bibfnamefont {R.~M.}\ \bibnamefont {Fernandes}},\
  }\href {\doibase 10.1103/PhysRevB.100.014512} {\bibfield  {journal} {\bibinfo
   {journal} {Physical Review B}\ }\textbf {\bibinfo {volume} {100}},\ \bibinfo
  {pages} {014512} (\bibinfo {year} {2019})}\BibitemShut {NoStop}%
\bibitem [{\citenamefont {Ruiz}\ \emph {et~al.}(2019)\citenamefont {Ruiz},
  \citenamefont {Wang}, \citenamefont {Moritz}, \citenamefont {Baum},
  \citenamefont {Hackl},\ and\ \citenamefont {Devereaux}}]{Ruiz.2019}%
  \BibitemOpen
  \bibfield  {author} {\bibinfo {author} {\bibfnamefont {H.}~\bibnamefont
  {Ruiz}}, \bibinfo {author} {\bibfnamefont {Y.}~\bibnamefont {Wang}}, \bibinfo
  {author} {\bibfnamefont {B.}~\bibnamefont {Moritz}}, \bibinfo {author}
  {\bibfnamefont {A.}~\bibnamefont {Baum}}, \bibinfo {author} {\bibfnamefont
  {R.}~\bibnamefont {Hackl}}, \ and\ \bibinfo {author} {\bibfnamefont {T.~P.}\
  \bibnamefont {Devereaux}},\ }\href {\doibase 10.1103/PhysRevB.99.125130}
  {\bibfield  {journal} {\bibinfo  {journal} {Physical Review B}\ }\textbf
  {\bibinfo {volume} {99}},\ \bibinfo {pages} {125130} (\bibinfo {year}
  {2019})}\BibitemShut {NoStop}%
\bibitem [{\citenamefont {Busemeyer}\ \emph {et~al.}(2016)\citenamefont
  {Busemeyer}, \citenamefont {Dagrada}, \citenamefont {Sorella}, \citenamefont
  {Casula},\ and\ \citenamefont {Wagner}}]{Busemeyer.2016}%
  \BibitemOpen
  \bibfield  {author} {\bibinfo {author} {\bibfnamefont {B.}~\bibnamefont
  {Busemeyer}}, \bibinfo {author} {\bibfnamefont {M.}~\bibnamefont {Dagrada}},
  \bibinfo {author} {\bibfnamefont {S.}~\bibnamefont {Sorella}}, \bibinfo
  {author} {\bibfnamefont {M.}~\bibnamefont {Casula}}, \ and\ \bibinfo {author}
  {\bibfnamefont {L.~K.}\ \bibnamefont {Wagner}},\ }\href {\doibase
  10.1103/PhysRevB.94.035108} {\bibfield  {journal} {\bibinfo  {journal}
  {Physical Review B}\ }\textbf {\bibinfo {volume} {94}},\ \bibinfo {pages}
  {035108} (\bibinfo {year} {2016})}\BibitemShut {NoStop}%
\bibitem [{\citenamefont {Baum}\ \emph {et~al.}(2019)\citenamefont {Baum},
  \citenamefont {Ruiz}, \citenamefont {Lazarevi{\'c}}, \citenamefont {Wang},
  \citenamefont {B{\"o}hm}, \citenamefont {{Hosseinian Ahangharnejhad}},
  \citenamefont {Adelmann}, \citenamefont {Wolf}, \citenamefont {Popovi{\'c}},
  \citenamefont {Moritz}, \citenamefont {Devereaux},\ and\ \citenamefont
  {Hackl}}]{Baum.2019}%
  \BibitemOpen
  \bibfield  {author} {\bibinfo {author} {\bibfnamefont {A.}~\bibnamefont
  {Baum}}, \bibinfo {author} {\bibfnamefont {H.~N.}\ \bibnamefont {Ruiz}},
  \bibinfo {author} {\bibfnamefont {N.}~\bibnamefont {Lazarevi{\'c}}}, \bibinfo
  {author} {\bibfnamefont {Y.}~\bibnamefont {Wang}}, \bibinfo {author}
  {\bibfnamefont {T.}~\bibnamefont {B{\"o}hm}}, \bibinfo {author}
  {\bibfnamefont {R.}~\bibnamefont {{Hosseinian Ahangharnejhad}}}, \bibinfo
  {author} {\bibfnamefont {P.}~\bibnamefont {Adelmann}}, \bibinfo {author}
  {\bibfnamefont {T.}~\bibnamefont {Wolf}}, \bibinfo {author} {\bibfnamefont
  {Z.~V.}\ \bibnamefont {Popovi{\'c}}}, \bibinfo {author} {\bibfnamefont
  {B.}~\bibnamefont {Moritz}}, \bibinfo {author} {\bibfnamefont {T.~P.}\
  \bibnamefont {Devereaux}}, \ and\ \bibinfo {author} {\bibfnamefont
  {R.}~\bibnamefont {Hackl}},\ }\href {\doibase 10.1038/s42005-019-0107-y}
  {\bibfield  {journal} {\bibinfo  {journal} {Communications Physics}\ }\textbf
  {\bibinfo {volume} {2}},\ \bibinfo {pages} {932} (\bibinfo {year}
  {2019})}\BibitemShut {NoStop}%
\bibitem [{\citenamefont {B{\"o}hmer}\ \emph {et~al.}(2019)\citenamefont
  {B{\"o}hmer}, \citenamefont {Kothapalli}, \citenamefont {Jayasekara},
  \citenamefont {Wilde}, \citenamefont {Li}, \citenamefont {Sapkota},
  \citenamefont {Ueland}, \citenamefont {Das}, \citenamefont {Xiao},
  \citenamefont {Bi}, \citenamefont {Zhao}, \citenamefont {Alp}, \citenamefont
  {Bud'ko}, \citenamefont {Canfield}, \citenamefont {Goldman},\ and\
  \citenamefont {Kreyssig}}]{Bohmer.2019}%
  \BibitemOpen
  \bibfield  {author} {\bibinfo {author} {\bibfnamefont {A.~E.}\ \bibnamefont
  {B{\"o}hmer}}, \bibinfo {author} {\bibfnamefont {K.}~\bibnamefont
  {Kothapalli}}, \bibinfo {author} {\bibfnamefont {W.~T.}\ \bibnamefont
  {Jayasekara}}, \bibinfo {author} {\bibfnamefont {J.~M.}\ \bibnamefont
  {Wilde}}, \bibinfo {author} {\bibfnamefont {B.}~\bibnamefont {Li}}, \bibinfo
  {author} {\bibfnamefont {A.}~\bibnamefont {Sapkota}}, \bibinfo {author}
  {\bibfnamefont {B.~G.}\ \bibnamefont {Ueland}}, \bibinfo {author}
  {\bibfnamefont {P.}~\bibnamefont {Das}}, \bibinfo {author} {\bibfnamefont
  {Y.}~\bibnamefont {Xiao}}, \bibinfo {author} {\bibfnamefont {W.}~\bibnamefont
  {Bi}}, \bibinfo {author} {\bibfnamefont {J.}~\bibnamefont {Zhao}}, \bibinfo
  {author} {\bibfnamefont {E.~E.}\ \bibnamefont {Alp}}, \bibinfo {author}
  {\bibfnamefont {S.~L.}\ \bibnamefont {Bud'ko}}, \bibinfo {author}
  {\bibfnamefont {P.~C.}\ \bibnamefont {Canfield}}, \bibinfo {author}
  {\bibfnamefont {A.~I.}\ \bibnamefont {Goldman}}, \ and\ \bibinfo {author}
  {\bibfnamefont {A.}~\bibnamefont {Kreyssig}},\ }\href {\doibase
  10.1103/PhysRevB.100.064515} {\bibfield  {journal} {\bibinfo  {journal}
  {Physical Review B}\ }\textbf {\bibinfo {volume} {100}},\ \bibinfo {pages}
  {064515} (\bibinfo {year} {2019})}\BibitemShut {NoStop}%
\bibitem [{\citenamefont {Wang}\ \emph
  {et~al.}(2016{\natexlab{a}})\citenamefont {Wang}, \citenamefont {Shen},
  \citenamefont {Pan}, \citenamefont {Zhang}, \citenamefont {Ikeuchi},
  \citenamefont {Iida}, \citenamefont {Christianson}, \citenamefont {Walker},
  \citenamefont {Adroja}, \citenamefont {Abdel-Hafiez}, \citenamefont {Chen},
  \citenamefont {Chareev}, \citenamefont {Vasiliev},\ and\ \citenamefont
  {Zhao}}]{Wang.2016c}%
  \BibitemOpen
  \bibfield  {author} {\bibinfo {author} {\bibfnamefont {Q.}~\bibnamefont
  {Wang}}, \bibinfo {author} {\bibfnamefont {Y.}~\bibnamefont {Shen}}, \bibinfo
  {author} {\bibfnamefont {B.}~\bibnamefont {Pan}}, \bibinfo {author}
  {\bibfnamefont {X.}~\bibnamefont {Zhang}}, \bibinfo {author} {\bibfnamefont
  {K.}~\bibnamefont {Ikeuchi}}, \bibinfo {author} {\bibfnamefont
  {K.}~\bibnamefont {Iida}}, \bibinfo {author} {\bibfnamefont {A.~D.}\
  \bibnamefont {Christianson}}, \bibinfo {author} {\bibfnamefont {H.~C.}\
  \bibnamefont {Walker}}, \bibinfo {author} {\bibfnamefont {D.~T.}\
  \bibnamefont {Adroja}}, \bibinfo {author} {\bibfnamefont {M.}~\bibnamefont
  {Abdel-Hafiez}}, \bibinfo {author} {\bibfnamefont {X.}~\bibnamefont {Chen}},
  \bibinfo {author} {\bibfnamefont {D.~A.}\ \bibnamefont {Chareev}}, \bibinfo
  {author} {\bibfnamefont {A.~N.}\ \bibnamefont {Vasiliev}}, \ and\ \bibinfo
  {author} {\bibfnamefont {J.}~\bibnamefont {Zhao}},\ }\href {\doibase
  10.1038/ncomms12182} {\bibfield  {journal} {\bibinfo  {journal} {Nature
  communications}\ }\textbf {\bibinfo {volume} {7}},\ \bibinfo {pages} {12182}
  (\bibinfo {year} {2016}{\natexlab{a}})}\BibitemShut {NoStop}%
\bibitem [{\citenamefont {Wang}\ \emph
  {et~al.}(2016{\natexlab{b}})\citenamefont {Wang}, \citenamefont {Shen},
  \citenamefont {Pan}, \citenamefont {Hao}, \citenamefont {Ma}, \citenamefont
  {Zhou}, \citenamefont {Steffens}, \citenamefont {Schmalzl}, \citenamefont
  {Forrest}, \citenamefont {Abdel-Hafiez}, \citenamefont {Chen}, \citenamefont
  {Chareev}, \citenamefont {Vasiliev}, \citenamefont {Bourges}, \citenamefont
  {Sidis}, \citenamefont {Cao},\ and\ \citenamefont {Zhao}}]{Wang.2016d}%
  \BibitemOpen
  \bibfield  {author} {\bibinfo {author} {\bibfnamefont {Q.}~\bibnamefont
  {Wang}}, \bibinfo {author} {\bibfnamefont {Y.}~\bibnamefont {Shen}}, \bibinfo
  {author} {\bibfnamefont {B.}~\bibnamefont {Pan}}, \bibinfo {author}
  {\bibfnamefont {Y.}~\bibnamefont {Hao}}, \bibinfo {author} {\bibfnamefont
  {M.}~\bibnamefont {Ma}}, \bibinfo {author} {\bibfnamefont {F.}~\bibnamefont
  {Zhou}}, \bibinfo {author} {\bibfnamefont {P.}~\bibnamefont {Steffens}},
  \bibinfo {author} {\bibfnamefont {K.}~\bibnamefont {Schmalzl}}, \bibinfo
  {author} {\bibfnamefont {T.~R.}\ \bibnamefont {Forrest}}, \bibinfo {author}
  {\bibfnamefont {M.}~\bibnamefont {Abdel-Hafiez}}, \bibinfo {author}
  {\bibfnamefont {X.}~\bibnamefont {Chen}}, \bibinfo {author} {\bibfnamefont
  {D.~A.}\ \bibnamefont {Chareev}}, \bibinfo {author} {\bibfnamefont {A.~N.}\
  \bibnamefont {Vasiliev}}, \bibinfo {author} {\bibfnamefont {P.}~\bibnamefont
  {Bourges}}, \bibinfo {author} {\bibfnamefont {Y.}~\bibnamefont {Sidis}},
  \bibinfo {author} {\bibfnamefont {H.}~\bibnamefont {Cao}}, \ and\ \bibinfo
  {author} {\bibfnamefont {J.}~\bibnamefont {Zhao}},\ }\href {\doibase
  10.1038/nmat4492} {\bibfield  {journal} {\bibinfo  {journal} {Nature
  materials}\ }\textbf {\bibinfo {volume} {15}},\ \bibinfo {pages} {159}
  (\bibinfo {year} {2016}{\natexlab{b}})}\BibitemShut {NoStop}%
\bibitem [{\citenamefont {Walle}\ and\ \citenamefont
  {Ceder}(2002)}]{Walle.2002}%
  \BibitemOpen
  \bibfield  {author} {\bibinfo {author} {\bibfnamefont {A.}~\bibnamefont
  {Walle}}\ and\ \bibinfo {author} {\bibfnamefont {G.}~\bibnamefont {Ceder}},\
  }\href {\doibase 10.1361/105497102770331596} {\bibfield  {journal} {\bibinfo
  {journal} {Journal of Phase Equilibria}\ }\textbf {\bibinfo {volume} {23}},\
  \bibinfo {pages} {348} (\bibinfo {year} {2002})}\BibitemShut {NoStop}%
\bibitem [{\citenamefont {Guterding}\ \emph
  {et~al.}(2017{\natexlab{b}})\citenamefont {Guterding}, \citenamefont
  {Jeschke},\ and\ \citenamefont {Valent{\'i}}}]{Guterding.2017b}%
  \BibitemOpen
  \bibfield  {author} {\bibinfo {author} {\bibfnamefont {D.}~\bibnamefont
  {Guterding}}, \bibinfo {author} {\bibfnamefont {H.~O.}\ \bibnamefont
  {Jeschke}}, \ and\ \bibinfo {author} {\bibfnamefont {R.}~\bibnamefont
  {Valent{\'i}}},\ }\href {\doibase 10.1103/PhysRevB.96.125107} {\bibfield
  {journal} {\bibinfo  {journal} {Physical Review B}\ }\textbf {\bibinfo
  {volume} {96}},\ \bibinfo {pages} {125107} (\bibinfo {year}
  {2017}{\natexlab{b}})}\BibitemShut {NoStop}%
\bibitem [{\citenamefont {Ricci}\ and\ \citenamefont
  {Profeta}(2013)}]{Ricci.2013}%
  \BibitemOpen
  \bibfield  {author} {\bibinfo {author} {\bibfnamefont {F.}~\bibnamefont
  {Ricci}}\ and\ \bibinfo {author} {\bibfnamefont {G.}~\bibnamefont
  {Profeta}},\ }\href {\doibase 10.1103/PhysRevB.87.184105} {\bibfield
  {journal} {\bibinfo  {journal} {Physical Review B}\ }\textbf {\bibinfo
  {volume} {87}},\ \bibinfo {pages} {184105} (\bibinfo {year}
  {2013})}\BibitemShut {NoStop}%
\bibitem [{\citenamefont {Grimme}(2006)}]{Grimme.2006}%
  \BibitemOpen
  \bibfield  {author} {\bibinfo {author} {\bibfnamefont {S.}~\bibnamefont
  {Grimme}},\ }\href {\doibase 10.1002/jcc.20495} {\bibfield  {journal}
  {\bibinfo  {journal} {Journal of Computational Chemistry}\ }\textbf {\bibinfo
  {volume} {27}},\ \bibinfo {pages} {1787} (\bibinfo {year}
  {2006})}\BibitemShut {NoStop}%
\bibitem [{\citenamefont {Tkatchenko}\ and\ \citenamefont
  {Scheffler}(2009)}]{Tkatchenko.2009}%
  \BibitemOpen
  \bibfield  {author} {\bibinfo {author} {\bibfnamefont {A.}~\bibnamefont
  {Tkatchenko}}\ and\ \bibinfo {author} {\bibfnamefont {M.}~\bibnamefont
  {Scheffler}},\ }\href {\doibase 10.1103/PhysRevLett.102.073005} {\bibfield
  {journal} {\bibinfo  {journal} {Physical Review Letters}\ }\textbf {\bibinfo
  {volume} {102}},\ \bibinfo {pages} {073005} (\bibinfo {year}
  {2009})}\BibitemShut {NoStop}%
\bibitem [{\citenamefont {Bleskov}\ \emph {et~al.}(2016)\citenamefont
  {Bleskov}, \citenamefont {Hickel}, \citenamefont {Neugebauer},\ and\
  \citenamefont {Ruban}}]{Bleskov.2016}%
  \BibitemOpen
  \bibfield  {author} {\bibinfo {author} {\bibfnamefont {I.}~\bibnamefont
  {Bleskov}}, \bibinfo {author} {\bibfnamefont {T.}~\bibnamefont {Hickel}},
  \bibinfo {author} {\bibfnamefont {J.}~\bibnamefont {Neugebauer}}, \ and\
  \bibinfo {author} {\bibfnamefont {A.}~\bibnamefont {Ruban}},\ }\href
  {\doibase 10.1103/PhysRevB.93.214115} {\bibfield  {journal} {\bibinfo
  {journal} {Physical Review B}\ }\textbf {\bibinfo {volume} {93}},\ \bibinfo
  {pages} {214115} (\bibinfo {year} {2016})}\BibitemShut {NoStop}%
\bibitem [{\citenamefont {Wu}\ and\ \citenamefont {Yang}(2002)}]{Wu.2002}%
  \BibitemOpen
  \bibfield  {author} {\bibinfo {author} {\bibfnamefont {Q.}~\bibnamefont
  {Wu}}\ and\ \bibinfo {author} {\bibfnamefont {W.}~\bibnamefont {Yang}},\
  }\href {\doibase 10.1063/1.1424928} {\bibfield  {journal} {\bibinfo
  {journal} {The Journal of Chemical Physics}\ }\textbf {\bibinfo {volume}
  {116}},\ \bibinfo {pages} {515} (\bibinfo {year} {2002})}\BibitemShut
  {NoStop}%
\bibitem [{\citenamefont {Chu}\ and\ \citenamefont
  {Dalgarno}(2004)}]{Chu.2004}%
  \BibitemOpen
  \bibfield  {author} {\bibinfo {author} {\bibfnamefont {X.}~\bibnamefont
  {Chu}}\ and\ \bibinfo {author} {\bibfnamefont {A.}~\bibnamefont {Dalgarno}},\
  }\href {\doibase 10.1063/1.1779576} {\bibfield  {journal} {\bibinfo
  {journal} {The Journal of Chemical Physics}\ }\textbf {\bibinfo {volume}
  {121}},\ \bibinfo {pages} {4083} (\bibinfo {year} {2004})}\BibitemShut
  {NoStop}%
\bibitem [{\citenamefont {Tang}(1969)}]{Tang.1969}%
  \BibitemOpen
  \bibfield  {author} {\bibinfo {author} {\bibfnamefont {K.~T.}\ \bibnamefont
  {Tang}},\ }\href {\doibase 10.1103/PhysRev.177.108} {\bibfield  {journal}
  {\bibinfo  {journal} {Physical Review}\ }\textbf {\bibinfo {volume} {177}},\
  \bibinfo {pages} {108} (\bibinfo {year} {1969})}\BibitemShut {NoStop}%
\bibitem [{\citenamefont {Hirschfeld}(1977)}]{Hirschfeld.1977}%
  \BibitemOpen
  \bibfield  {author} {\bibinfo {author} {\bibfnamefont {F.~L.}\ \bibnamefont
  {Hirschfeld}},\ }\href {\doibase 10.1007/BF00549096} {\bibfield  {journal}
  {\bibinfo  {journal} {Theoret. Claim. Acta (Berl.)}\ }\textbf {\bibinfo
  {volume} {44}},\ \bibinfo {pages} {129} (\bibinfo {year} {1977})}\BibitemShut
  {NoStop}%
\bibitem [{\citenamefont {Kresse}\ and\ \citenamefont
  {Hafner}(1993)}]{Kresse.1993}%
  \BibitemOpen
  \bibfield  {author} {\bibinfo {author} {\bibfnamefont {G.}~\bibnamefont
  {Kresse}}\ and\ \bibinfo {author} {\bibfnamefont {J.}~\bibnamefont
  {Hafner}},\ }\href {\doibase 10.1103/PhysRevB.47.558} {\bibfield  {journal}
  {\bibinfo  {journal} {Physical Review B}\ }\textbf {\bibinfo {volume} {47}},\
  \bibinfo {pages} {558} (\bibinfo {year} {1993})}\BibitemShut {NoStop}%
\bibitem [{\citenamefont {Kresse}\ and\ \citenamefont
  {Furthm{\"u}ller}(1996{\natexlab{a}})}]{Kresse.1996}%
  \BibitemOpen
  \bibfield  {author} {\bibinfo {author} {\bibfnamefont {G.}~\bibnamefont
  {Kresse}}\ and\ \bibinfo {author} {\bibfnamefont {J.}~\bibnamefont
  {Furthm{\"u}ller}},\ }\href {\doibase 10.1103/PhysRevB.54.11169} {\bibfield
  {journal} {\bibinfo  {journal} {Physical Review B}\ }\textbf {\bibinfo
  {volume} {54}},\ \bibinfo {pages} {11169} (\bibinfo {year}
  {1996}{\natexlab{a}})}\BibitemShut {NoStop}%
\bibitem [{\citenamefont {Kresse}\ and\ \citenamefont
  {Furthm{\"u}ller}(1996{\natexlab{b}})}]{Kresse.1996b}%
  \BibitemOpen
  \bibfield  {author} {\bibinfo {author} {\bibfnamefont {G.}~\bibnamefont
  {Kresse}}\ and\ \bibinfo {author} {\bibfnamefont {J.}~\bibnamefont
  {Furthm{\"u}ller}},\ }\href {\doibase 10.1016/0927-0256(96)00008-0}
  {\bibfield  {journal} {\bibinfo  {journal} {Computational Materials Science}\
  }\textbf {\bibinfo {volume} {6}},\ \bibinfo {pages} {15} (\bibinfo {year}
  {1996}{\natexlab{b}})}\BibitemShut {NoStop}%
\bibitem [{\citenamefont {Bl{\"o}chl}(1994)}]{Blochl.1994}%
  \BibitemOpen
  \bibfield  {author} {\bibinfo {author} {\bibfnamefont {P.~E.}\ \bibnamefont
  {Bl{\"o}chl}},\ }\href {\doibase 10.1103/PhysRevB.50.17953} {\bibfield
  {journal} {\bibinfo  {journal} {Physical Review B}\ }\textbf {\bibinfo
  {volume} {50}},\ \bibinfo {pages} {17953} (\bibinfo {year}
  {1994})}\BibitemShut {NoStop}%
\bibitem [{\citenamefont {Perdew}\ \emph {et~al.}(1996)\citenamefont {Perdew},
  \citenamefont {Burke},\ and\ \citenamefont {Ernzerhof}}]{Perdew.1996}%
  \BibitemOpen
  \bibfield  {author} {\bibinfo {author} {\bibnamefont {Perdew}}, \bibinfo
  {author} {\bibnamefont {Burke}}, \ and\ \bibinfo {author} {\bibnamefont
  {Ernzerhof}},\ }\href {\doibase 10.1103/PhysRevLett.77.3865} {\bibfield
  {journal} {\bibinfo  {journal} {Physical Review Letters}\ }\textbf {\bibinfo
  {volume} {77}},\ \bibinfo {pages} {3865} (\bibinfo {year}
  {1996})}\BibitemShut {NoStop}%
\bibitem [{\citenamefont {Monkhorst}\ and\ \citenamefont
  {Pack}(1976)}]{Monkhorst.1976}%
  \BibitemOpen
  \bibfield  {author} {\bibinfo {author} {\bibfnamefont {H.~J.}\ \bibnamefont
  {Monkhorst}}\ and\ \bibinfo {author} {\bibfnamefont {J.~D.}\ \bibnamefont
  {Pack}},\ }\href {\doibase 10.1103/PhysRevB.13.5188} {\bibfield  {journal}
  {\bibinfo  {journal} {Physical Review B}\ }\textbf {\bibinfo {volume} {13}},\
  \bibinfo {pages} {5188} (\bibinfo {year} {1976})}\BibitemShut {NoStop}%
\bibitem [{\citenamefont {Methfessel}\ and\ \citenamefont
  {Paxton}(1989)}]{Methfessel.1989}%
  \BibitemOpen
  \bibfield  {author} {\bibinfo {author} {\bibnamefont {Methfessel}}\ and\
  \bibinfo {author} {\bibnamefont {Paxton}},\ }\href {\doibase
  10.1103/physrevb.40.3616} {\bibfield  {journal} {\bibinfo  {journal}
  {Physical Review B}\ }\textbf {\bibinfo {volume} {40}},\ \bibinfo {pages}
  {3616} (\bibinfo {year} {1989})}\BibitemShut {NoStop}%
\bibitem [{\citenamefont {Janssen}\ \emph {et~al.}(2019)\citenamefont
  {Janssen}, \citenamefont {Surendralal}, \citenamefont {Lysogorskiy},
  \citenamefont {Todorova}, \citenamefont {Hickel}, \citenamefont {Drautz},\
  and\ \citenamefont {Neugebauer}}]{Janssen.2019}%
  \BibitemOpen
  \bibfield  {author} {\bibinfo {author} {\bibfnamefont {J.}~\bibnamefont
  {Janssen}}, \bibinfo {author} {\bibfnamefont {S.}~\bibnamefont
  {Surendralal}}, \bibinfo {author} {\bibfnamefont {Y.}~\bibnamefont
  {Lysogorskiy}}, \bibinfo {author} {\bibfnamefont {M.}~\bibnamefont
  {Todorova}}, \bibinfo {author} {\bibfnamefont {T.}~\bibnamefont {Hickel}},
  \bibinfo {author} {\bibfnamefont {R.}~\bibnamefont {Drautz}}, \ and\ \bibinfo
  {author} {\bibfnamefont {J.}~\bibnamefont {Neugebauer}},\ }\href {\doibase
  10.1016/j.commatsci.2018.07.043} {\bibfield  {journal} {\bibinfo  {journal}
  {Computational Materials Science}\ }\textbf {\bibinfo {volume} {163}},\
  \bibinfo {pages} {24} (\bibinfo {year} {2019})}\BibitemShut {NoStop}%
\bibitem [{\citenamefont {McQueen}\ \emph
  {et~al.}(2009{\natexlab{b}})\citenamefont {McQueen}, \citenamefont {Huang},
  \citenamefont {Ksenofontov}, \citenamefont {Felser}, \citenamefont {Xu},
  \citenamefont {Zandbergen}, \citenamefont {Hor}, \citenamefont {Allred},
  \citenamefont {Williams}, \citenamefont {Qu}, \citenamefont {Checkelsky},
  \citenamefont {Ong},\ and\ \citenamefont {Cava}}]{McQueen.2009}%
  \BibitemOpen
  \bibfield  {author} {\bibinfo {author} {\bibfnamefont {T.~M.}\ \bibnamefont
  {McQueen}}, \bibinfo {author} {\bibfnamefont {Q.}~\bibnamefont {Huang}},
  \bibinfo {author} {\bibfnamefont {V.}~\bibnamefont {Ksenofontov}}, \bibinfo
  {author} {\bibfnamefont {C.}~\bibnamefont {Felser}}, \bibinfo {author}
  {\bibfnamefont {Q.}~\bibnamefont {Xu}}, \bibinfo {author} {\bibfnamefont
  {H.}~\bibnamefont {Zandbergen}}, \bibinfo {author} {\bibfnamefont {Y.~S.}\
  \bibnamefont {Hor}}, \bibinfo {author} {\bibfnamefont {J.}~\bibnamefont
  {Allred}}, \bibinfo {author} {\bibfnamefont {A.~J.}\ \bibnamefont
  {Williams}}, \bibinfo {author} {\bibfnamefont {D.}~\bibnamefont {Qu}},
  \bibinfo {author} {\bibfnamefont {J.}~\bibnamefont {Checkelsky}}, \bibinfo
  {author} {\bibfnamefont {N.~P.}\ \bibnamefont {Ong}}, \ and\ \bibinfo
  {author} {\bibfnamefont {R.~J.}\ \bibnamefont {Cava}},\ }\href {\doibase
  10.1103/PhysRevB.79.014522} {\bibfield  {journal} {\bibinfo  {journal}
  {Physical Review B}\ }\textbf {\bibinfo {volume} {79}},\ \bibinfo {pages}
  {014522} (\bibinfo {year} {2009}{\natexlab{b}})}\BibitemShut {NoStop}%
\bibitem [{\citenamefont {Vinet}\ \emph {et~al.}(1987)\citenamefont {Vinet},
  \citenamefont {Smith}, \citenamefont {Ferrante},\ and\ \citenamefont
  {Rose}}]{Vinet.1987}%
  \BibitemOpen
  \bibfield  {author} {\bibinfo {author} {\bibfnamefont {P.}~\bibnamefont
  {Vinet}}, \bibinfo {author} {\bibfnamefont {J.~R.}\ \bibnamefont {Smith}},
  \bibinfo {author} {\bibfnamefont {J.}~\bibnamefont {Ferrante}}, \ and\
  \bibinfo {author} {\bibfnamefont {J.~H.}\ \bibnamefont {Rose}},\ }\href
  {\doibase 10.1103/physrevb.35.1945} {\bibfield  {journal} {\bibinfo
  {journal} {Physical Review B}\ }\textbf {\bibinfo {volume} {35}},\ \bibinfo
  {pages} {1945} (\bibinfo {year} {1987})}\BibitemShut {NoStop}%
\bibitem [{\citenamefont {Gastiasoro}\ and\ \citenamefont
  {Andersen}(2015)}]{Gastiasoro.2015}%
  \BibitemOpen
  \bibfield  {author} {\bibinfo {author} {\bibfnamefont {M.~N.}\ \bibnamefont
  {Gastiasoro}}\ and\ \bibinfo {author} {\bibfnamefont {B.~M.}\ \bibnamefont
  {Andersen}},\ }\href {\doibase 10.1103/PhysRevB.92.140506} {\bibfield
  {journal} {\bibinfo  {journal} {Physical Review B}\ }\textbf {\bibinfo
  {volume} {92}},\ \bibinfo {pages} {140506} (\bibinfo {year}
  {2015})}\BibitemShut {NoStop}%
\bibitem [{\citenamefont {Millican}\ \emph {et~al.}(2009)\citenamefont
  {Millican}, \citenamefont {Phelan}, \citenamefont {Thomas}, \citenamefont
  {Le{\~a}o},\ and\ \citenamefont {Carpenter}}]{Millican.2009}%
  \BibitemOpen
  \bibfield  {author} {\bibinfo {author} {\bibfnamefont {J.~N.}\ \bibnamefont
  {Millican}}, \bibinfo {author} {\bibfnamefont {D.}~\bibnamefont {Phelan}},
  \bibinfo {author} {\bibfnamefont {E.~L.}\ \bibnamefont {Thomas}}, \bibinfo
  {author} {\bibfnamefont {J.~B.}\ \bibnamefont {Le{\~a}o}}, \ and\ \bibinfo
  {author} {\bibfnamefont {E.}~\bibnamefont {Carpenter}},\ }\href {\doibase
  10.1016/j.ssc.2009.02.011} {\bibfield  {journal} {\bibinfo  {journal} {Solid
  State Communications}\ }\textbf {\bibinfo {volume} {149}},\ \bibinfo {pages}
  {707} (\bibinfo {year} {2009})}\BibitemShut {NoStop}%
\bibitem [{\citenamefont {Margadonna}\ \emph {et~al.}(2009)\citenamefont
  {Margadonna}, \citenamefont {Takabayashi}, \citenamefont {Ohishi},
  \citenamefont {Mizuguchi}, \citenamefont {Takano}, \citenamefont {Kagayama},
  \citenamefont {Nakagawa}, \citenamefont {Takata},\ and\ \citenamefont
  {Prassides}}]{Margadonna.2009}%
  \BibitemOpen
  \bibfield  {author} {\bibinfo {author} {\bibfnamefont {S.}~\bibnamefont
  {Margadonna}}, \bibinfo {author} {\bibfnamefont {Y.}~\bibnamefont
  {Takabayashi}}, \bibinfo {author} {\bibfnamefont {Y.}~\bibnamefont {Ohishi}},
  \bibinfo {author} {\bibfnamefont {Y.}~\bibnamefont {Mizuguchi}}, \bibinfo
  {author} {\bibfnamefont {Y.}~\bibnamefont {Takano}}, \bibinfo {author}
  {\bibfnamefont {T.}~\bibnamefont {Kagayama}}, \bibinfo {author}
  {\bibfnamefont {T.}~\bibnamefont {Nakagawa}}, \bibinfo {author}
  {\bibfnamefont {M.}~\bibnamefont {Takata}}, \ and\ \bibinfo {author}
  {\bibfnamefont {K.}~\bibnamefont {Prassides}},\ }\href {\doibase
  10.1103/PhysRevB.80.064506} {\bibfield  {journal} {\bibinfo  {journal}
  {Physical Review B}\ }\textbf {\bibinfo {volume} {80}},\ \bibinfo {pages}
  {064506} (\bibinfo {year} {2009})}\BibitemShut {NoStop}%
\bibitem [{\citenamefont {Koz}\ \emph {et~al.}(2013)\citenamefont {Koz},
  \citenamefont {R{\"o}{\ss}ler}, \citenamefont {Tsirlin}, \citenamefont
  {Wirth},\ and\ \citenamefont {Schwarz}}]{Koz.2013}%
  \BibitemOpen
  \bibfield  {author} {\bibinfo {author} {\bibfnamefont {C.}~\bibnamefont
  {Koz}}, \bibinfo {author} {\bibfnamefont {S.}~\bibnamefont {R{\"o}{\ss}ler}},
  \bibinfo {author} {\bibfnamefont {A.~A.}\ \bibnamefont {Tsirlin}}, \bibinfo
  {author} {\bibfnamefont {S.}~\bibnamefont {Wirth}}, \ and\ \bibinfo {author}
  {\bibfnamefont {U.}~\bibnamefont {Schwarz}},\ }\href {\doibase
  10.1103/PhysRevB.88.094509} {\bibfield  {journal} {\bibinfo  {journal}
  {Physical Review B}\ }\textbf {\bibinfo {volume} {88}},\ \bibinfo {pages}
  {094509} (\bibinfo {year} {2013})}\BibitemShut {NoStop}%
\bibitem [{\citenamefont {Allen}\ \emph {et~al.}(2010)\citenamefont {Allen},
  \citenamefont {Tung},\ and\ \citenamefont {Kaner}}]{Allen.2010}%
  \BibitemOpen
  \bibfield  {author} {\bibinfo {author} {\bibfnamefont {M.~J.}\ \bibnamefont
  {Allen}}, \bibinfo {author} {\bibfnamefont {V.~C.}\ \bibnamefont {Tung}}, \
  and\ \bibinfo {author} {\bibfnamefont {R.~B.}\ \bibnamefont {Kaner}},\ }\href
  {\doibase 10.1021/cr900070d} {\bibfield  {journal} {\bibinfo  {journal}
  {Chemical reviews}\ }\textbf {\bibinfo {volume} {110}},\ \bibinfo {pages}
  {132} (\bibinfo {year} {2010})}\BibitemShut {NoStop}%
\bibitem [{\citenamefont {Grabowski}\ \emph {et~al.}(2007)\citenamefont
  {Grabowski}, \citenamefont {Hickel},\ and\ \citenamefont
  {Neugebauer}}]{Grabowski2007}%
  \BibitemOpen
  \bibfield  {author} {\bibinfo {author} {\bibfnamefont {B.}~\bibnamefont
  {Grabowski}}, \bibinfo {author} {\bibfnamefont {T.}~\bibnamefont {Hickel}}, \
  and\ \bibinfo {author} {\bibfnamefont {J.}~\bibnamefont {Neugebauer}},\
  }\href {\doibase 10.1103/PhysRevB.76.024309} {\bibfield  {journal} {\bibinfo
  {journal} {Phys. Rev. B}\ }\textbf {\bibinfo {volume} {76}},\ \bibinfo
  {pages} {024309} (\bibinfo {year} {2007})}\BibitemShut {NoStop}%
\bibitem [{\citenamefont {Watson}\ \emph {et~al.}(2015)\citenamefont {Watson},
  \citenamefont {Kim}, \citenamefont {Haghighirad}, \citenamefont {Davies},
  \citenamefont {McCollam}, \citenamefont {Narayanan}, \citenamefont {Blake},
  \citenamefont {Chen}, \citenamefont {Ghannadzadeh}, \citenamefont
  {Schofield}, \citenamefont {Hoesch}, \citenamefont {Meingast}, \citenamefont
  {Wolf},\ and\ \citenamefont {Coldea}}]{Watson.2015}%
  \BibitemOpen
  \bibfield  {author} {\bibinfo {author} {\bibfnamefont {M.~D.}\ \bibnamefont
  {Watson}}, \bibinfo {author} {\bibfnamefont {T.~K.}\ \bibnamefont {Kim}},
  \bibinfo {author} {\bibfnamefont {A.~A.}\ \bibnamefont {Haghighirad}},
  \bibinfo {author} {\bibfnamefont {N.~R.}\ \bibnamefont {Davies}}, \bibinfo
  {author} {\bibfnamefont {A.}~\bibnamefont {McCollam}}, \bibinfo {author}
  {\bibfnamefont {A.}~\bibnamefont {Narayanan}}, \bibinfo {author}
  {\bibfnamefont {S.~F.}\ \bibnamefont {Blake}}, \bibinfo {author}
  {\bibfnamefont {Y.~L.}\ \bibnamefont {Chen}}, \bibinfo {author}
  {\bibfnamefont {S.}~\bibnamefont {Ghannadzadeh}}, \bibinfo {author}
  {\bibfnamefont {A.~J.}\ \bibnamefont {Schofield}}, \bibinfo {author}
  {\bibfnamefont {M.}~\bibnamefont {Hoesch}}, \bibinfo {author} {\bibfnamefont
  {C.}~\bibnamefont {Meingast}}, \bibinfo {author} {\bibfnamefont
  {T.}~\bibnamefont {Wolf}}, \ and\ \bibinfo {author} {\bibfnamefont {A.~I.}\
  \bibnamefont {Coldea}},\ }\href {\doibase 10.1103/PhysRevB.91.155106}
  {\bibfield  {journal} {\bibinfo  {journal} {Physical Review B}\ }\textbf
  {\bibinfo {volume} {91}},\ \bibinfo {pages} {524} (\bibinfo {year}
  {2015})}\BibitemShut {NoStop}%
\bibitem [{\citenamefont {Martiny}\ \emph {et~al.}(2019)\citenamefont
  {Martiny}, \citenamefont {Kreisel},\ and\ \citenamefont
  {Andersen}}]{Martiny.2019}%
  \BibitemOpen
  \bibfield  {author} {\bibinfo {author} {\bibfnamefont {J.~H.~J.}\
  \bibnamefont {Martiny}}, \bibinfo {author} {\bibfnamefont {A.}~\bibnamefont
  {Kreisel}}, \ and\ \bibinfo {author} {\bibfnamefont {B.~M.}\ \bibnamefont
  {Andersen}},\ }\href {\doibase 10.1103/PhysRevB.99.014509} {\bibfield
  {journal} {\bibinfo  {journal} {Physical Review B}\ }\textbf {\bibinfo
  {volume} {99}},\ \bibinfo {pages} {014509} (\bibinfo {year}
  {2019})}\BibitemShut {NoStop}%
\bibitem [{\citenamefont {Zhang}\ \emph {et~al.}(2018)\citenamefont {Zhang},
  \citenamefont {Wang}, \citenamefont {Wu}, \citenamefont {Yaji}, \citenamefont
  {Ishida}, \citenamefont {Kohama}, \citenamefont {Dai}, \citenamefont {Sun},
  \citenamefont {Bareille}, \citenamefont {Kuroda}, \citenamefont {Kondo},
  \citenamefont {Okazaki}, \citenamefont {Kindo}, \citenamefont {Wang},
  \citenamefont {Jin}, \citenamefont {Hu}, \citenamefont {Thomale},
  \citenamefont {Sumida}, \citenamefont {Wu}, \citenamefont {Miyamoto},
  \citenamefont {Okuda}, \citenamefont {Ding}, \citenamefont {Gu},
  \citenamefont {Tamegai}, \citenamefont {Kawakami}, \citenamefont {Sato},\
  and\ \citenamefont {Shin}}]{Zhang.2018d}%
  \BibitemOpen
  \bibfield  {author} {\bibinfo {author} {\bibfnamefont {P.}~\bibnamefont
  {Zhang}}, \bibinfo {author} {\bibfnamefont {Z.}~\bibnamefont {Wang}},
  \bibinfo {author} {\bibfnamefont {X.}~\bibnamefont {Wu}}, \bibinfo {author}
  {\bibfnamefont {K.}~\bibnamefont {Yaji}}, \bibinfo {author} {\bibfnamefont
  {Y.}~\bibnamefont {Ishida}}, \bibinfo {author} {\bibfnamefont
  {Y.}~\bibnamefont {Kohama}}, \bibinfo {author} {\bibfnamefont
  {G.}~\bibnamefont {Dai}}, \bibinfo {author} {\bibfnamefont {Y.}~\bibnamefont
  {Sun}}, \bibinfo {author} {\bibfnamefont {C.}~\bibnamefont {Bareille}},
  \bibinfo {author} {\bibfnamefont {K.}~\bibnamefont {Kuroda}}, \bibinfo
  {author} {\bibfnamefont {T.}~\bibnamefont {Kondo}}, \bibinfo {author}
  {\bibfnamefont {K.}~\bibnamefont {Okazaki}}, \bibinfo {author} {\bibfnamefont
  {K.}~\bibnamefont {Kindo}}, \bibinfo {author} {\bibfnamefont
  {X.}~\bibnamefont {Wang}}, \bibinfo {author} {\bibfnamefont {C.}~\bibnamefont
  {Jin}}, \bibinfo {author} {\bibfnamefont {J.}~\bibnamefont {Hu}}, \bibinfo
  {author} {\bibfnamefont {R.}~\bibnamefont {Thomale}}, \bibinfo {author}
  {\bibfnamefont {K.}~\bibnamefont {Sumida}}, \bibinfo {author} {\bibfnamefont
  {S.}~\bibnamefont {Wu}}, \bibinfo {author} {\bibfnamefont {K.}~\bibnamefont
  {Miyamoto}}, \bibinfo {author} {\bibfnamefont {T.}~\bibnamefont {Okuda}},
  \bibinfo {author} {\bibfnamefont {H.}~\bibnamefont {Ding}}, \bibinfo {author}
  {\bibfnamefont {G.~D.}\ \bibnamefont {Gu}}, \bibinfo {author} {\bibfnamefont
  {T.}~\bibnamefont {Tamegai}}, \bibinfo {author} {\bibfnamefont
  {T.}~\bibnamefont {Kawakami}}, \bibinfo {author} {\bibfnamefont
  {M.}~\bibnamefont {Sato}}, \ and\ \bibinfo {author} {\bibfnamefont
  {S.}~\bibnamefont {Shin}},\ }\href {\doibase 10.1038/s41567-018-0280-z}
  {\bibfield  {journal} {\bibinfo  {journal} {Nature Physics}\ }\textbf
  {\bibinfo {volume} {130}},\ \bibinfo {pages} {3296} (\bibinfo {year}
  {2018})}\BibitemShut {NoStop}%
\end{thebibliography}%
%

\end{document}